\documentclass{aastex631}
\usepackage{xcolor}
\newcommand{\gtsim}{\raisebox{-1.0ex}{$\stackrel{\textstyle>}{\sim}$}}
\newcommand{\ltsim}{\raisebox{-1.0ex}{$\stackrel{\textstyle<}{\sim}$}}
\long\def\comment#1{}
\def\kms{km~s$^{-1}$}
\def\al{Alfv\'{e}n}

\def\yohkoh{{\sl Yohkoh}}

\def\hinode{{\sl Hinode}}

\def\p78{{\sl P78-1}}

\def\stereo{{\sl STEREO}}
\def\sdo{{\sl SDO}}

\def\heii{He~{\sc ii}}

\def\al{Alfv\'{e}n}

\def\kms{km~s$^{-1}$}

\begin{document}
%

\title{Another Look at Erupting Minifilaments at the Base of Solar X-Ray Polar Coronal 
``Standard'' and ``Blowout'' Jets}

\author{Alphonse C. Sterling}
\affiliation{NASA/Marshall Space Flight Center, Huntsville, AL 35812, USA}

\author{Ronald L. Moore} 
\affiliation{NASA/Marshall Space Flight Center, Huntsville, AL 35812, USA}
\affiliation{Center for Space Plasma and Aeronomic Research, \\
University of Alabama in Huntsville, Huntsville, AL 35805, USA}

\author{Navdeep K. Panesar} 
\affiliation{Lockheed Martin Solar and Astrophysics Laboratory, 3251 Hanover Street, Building 252, Palo Alto, CA 94304, USA}
\affiliation{Bay Area Environmental Research Institute, NASA Research Park, Moffett Field, CA 94035, USA}

\begin{abstract}

We examine 21 solar polar coronal jets that we identify in soft X-ray images obtained from the 
\hinode/X-ray telescope (XRT)\@. We identify 11 of these jets as blowout jets and four as standard jets 
(with six uncertain),
based on their X-ray-spire widths being respectively wide or narrow (compared to the jet's base) in the
XRT images.  From corresponding Extreme Ultraviolet (EUV) images from the Solar Dynamics Observatory's
(\sdo) Atmospheric Imaging Assembly (AIA), essentially all (at least 20 of 21) of the jets
are made by minifilament eruptions, consistent  with other recent studies. Here, we examine
the detailed nature of the erupting minifilaments (EMFs) in the jet bases.  Wide-spire (``blowout") 
jets often have ejective EMFs, but sometimes they instead have an EMF that is mostly confined to 
the jet's base rather than ejected.   We also demonstrate that narrow-spire (``standard") 
jets can have either a confined EMF, or a partially  confined EMF where some of the cool 
minifilament leaks into the jet's spire. 
Regarding EMF visibility: we find that in  some cases the minifilament is apparent in as few as 
one of the four EUV
channels we examined, being essentially invisible in the  other channels; thus it is necessary to examine
images from multiple EUV channels before concluding that a jet  does not have an EMF at its base.  The
size of the EMFs, measured  projected against the sky and early in their eruption, is $14''\pm 7''$, which
is within a factor of two of other measured sizes of coronal-jet EMFs.

 \end{abstract}

\keywords{Solar filament eruptions, solar corona, solar x-ray emission, solar extreme ultraviolet emission}

\section{Introduction}
\label{sec-introduction}

Solar coronal jets are transient features observed at coronal wavelengths, with spires  that
become long and narrow as they extend outward from the solar surface into the corona. The base
of the jets also become bright, and frequently the brightest part is located on one side of the
base \citep[a feature that we will call the {\it jet bright point}, JBP, 
following][]{sterling.et15}. Interest in jets increased dramatically with X-ray observations 
from \yohkoh\ \citep{shibata.et92}.  In
polar coronal holes, a typical coronal jet reportedly will reach a length of
$\sim$50{,}000\,km, have width  of $\sim$8000\,km, and live for $\sim$10\,min
\citep{savcheva.et07}.  In addition to appearing in coronal holes, they also occur in quiet
Sun, and at the periphery of active regions.  Following the \yohkoh\ studies, they have
also been observed in X-rays with the X-Ray Telescope (XRT) on \hinode\ \citep{cirtain.et07},
and they have also been widely  studied in EUV \citep[e.g.][]{nistico.et09} and at other
wavelengths.  Several reviews are now available describing historical and recent progress
\citep{shibata.et11,raouafi.et16,hinode.et19,innes.et16,shen21}.

An early idea for the origin of jets held that they resulted when new magnetic flux emerged
from below the photosphere and reconnected with encountered far-reaching coronal 
field  \citep{shibata.et92}. 
Various numerical simulation studies provided support for this {\it emerging flux model} for jet
production
\citep[e.g.,][]{yokoyama.et95,nishizuka.et08,moreno-insertis.et13,fang.et14}.   \yohkoh\ images
however only detected relatively hot coronal  emissions, and the magnetograms of the time were
mainly low-cadence (a few per day) from the ground.  It turns out that these limitations helped
to conceal other important properties of coronal jets.

Subsequent studies of jets over multiple EUV wavelengths, mainly from the Atmospheric Imaging
Assembly (AIA) on the Solar Dynamics Observatory (\sdo) spacecraft, along with  continuous
full-disk, high-cadence (45\,s) magnetograms from the Helioseismic and  Magnetic Imager (HMI),
also on \sdo, allowed for a new interpretation of the cause of  jets. Some studies of single 
jets indicated that a jet can result from eruption of a small-scale filament
\citep[e.g.,][]{shen.et12,hong.et14,adams.et14}.  Similar single-jet studies showed cases where a
jet came from a site where magnetic flux cancelation was occurring, rather than flux emergence
\citep[e.g.][]{shen.et12,hong.et14,young.et14a,young.et14b,adams.et14}.  \citet{sterling.et15} 
proposed what we call a {\it minifilament eruption model}, whereby essentially all jets result
from the eruption of such small-scale filaments, or minifilaments.  Here, we will refer to the
cool-material erupting minifilament by EMF\@.

The basic idea is presented in a schematic in Figure~2 of \citet{sterling.et15}; a slightly modified
version of the same schematic appears in Figure~1 of \citet{sterling.et18}, and we will refer
to labelings in that version (the same labelings are in Figure~\ref{f_schematic2}, below).  The 
schematic is based on \citet{sterling.et15}'s inference 
that, in a coronal hole consisting of a pervasive magnetic polarity (say,
negative) open field, there would be an opposite-polarity (positive) magnetic element embedded
at the jet location. (The same picture works for the pervasive field being a foot and leg of 
a large closed-loop coronal
field, as long as the coronal loop is much larger in length and foot diameter than the
jet-base bipole embedded inside the foot.  Here we discuss the situation in terms of open
field, both in order to simplify the discussion, and because the examples we deal with in this
paper are in polar coronal holes, where the coronal field is expected to be largely open.) This
embedded positive-polarity element would form an anemone-shaped magnetic structure
\citep{shibata.et07}, with lobes connecting the positive flux element to the coronal hole's
surrounding negative magnetic flux.  One lobe of this anemone would contain the minifilament
(the lobe connected by magnetic elements labeled M2 and M3 in the schematic shown in 
\citet{sterling.et18} (those same labels are included in the schematic of 
Figure~\ref{f_schematic2}(a), below), and
that minifilament (consisting of both the cool minifilament chromospheric material and the
magnetic field/flux rope enveloping and carrying that cool material) would erupt outward.  As
it does so, two reconnections ensue: One is {\it external reconnection} between the field
enveloping the EMF flux rope and the ambient open coronal field surrounding
the anemone.  That external reconnection would result in some of the field enveloping the
minifilament becoming open. Reconnection-heated material flowing out along that open field
would form the jet spire. Also, that  external reconnection would add new reconnection-heated
loops to the M1---M2 lobe, connecting the positive-polarity element with negative polarity on
the far side of that lobe. The other reconnection would be {\it internal reconnection} among the
legs of the enveloping field, occurring below the EMF's flux rope; this would form a closed 
heated arcade
of loops, the JBP, below the EMF, analogous to the flare arcade
forming below a typical large-scale erupting filament.   Thus, \citet{sterling.et15} argued that
coronal jets are  scaled-down versions of large-scale magnetic eruptions, that 
often make eruptions of typically sized filaments, typical solar flares, and coronal mass ejections
(CMEs), as in the case of the standard flare model
\citep[e.g.,][]{hirayama74,shibata.et95,moore.et01,chen11}.  In particular, in this view the
JBP forms under the EMF by internal reconnection in the erupting lobe,  not
by the external reconnection of the emerging lobe, by which the JBP was argued to form in the
emerging-flux model \citep{yokoyama.et95}.  \citet{sterling.et15} found in each of the 20 jets
they examined that the JBP formed at the location of the minifilament's eruption onset.

Other studies investigating the eruption mechanism of a selection of quiet Sun and  coronal hole jets 
have also found that they often result from eruptions of a minifilament.  This includes work by us
\citep[e.g.,][]{panesar.et16a,panesar.et17,panesar.et18a,mcglasson.et19}. Other groups have
also found similar results.  For example, \citet{kumar.et19} found that two-thirds of the equatorial coronal hole jets they studied included EMFs; for the 
remaining one-third of their sample they did not observe 
erupting cool-material minifilaments, but those events did exhibit other eruptive  signatures.  So \citet{kumar.et19} concluded that every jet of their study ``involved a filament-channel eruption." 
Other
workers have also reported EMFs in quiet Sun and coronal hole single-jet studies
\citep[e.g][]{shen.et17,doyle.et19,zhang.et21}.  Therefore, while we are unable to say whether all quiet
Sun and coronal hole jets result from EMFs, there is strong evidence that minifilament
eruptions at least often produce such jets.  In earlier studies, prior to AIA observations, often 
EMFs would not have been visible, due to poorer cadenced in the EUV \citep[although even in
that case some studies indicated jets resulted from small-scale eruptions; e.g.][]{nistico.et09}. 
Because the EMFs are not readily visible in X-ray images \citep{sterling.et15}, it is not surprising 
that the earliest 
studies in X-rays without access to quality EUV images \citep[e.g.][]{shimojo.et96}
would not have detected EMFs in X-ray images.

In subsequent studies, \citet{panesar.et16a} argued that the primary process leading to and
triggering the minifilament eruptions causing jets is magnetic flux cancelation.  Several 
earlier and later works support this view
\citep[e.g.,][]{hong.et11,huang.et12,shen.et12,young.et14a,adams.et14,sterling.et17,panesar.et17,panesar.et18a,mcglasson.et19,muglach21}. 
\citet{kumar.et19}, however, argue that shearing and/or field rotation may be
more important than cancelation in many jets.

A series of numerical simulations
\citep{wyper.et17,wyper.et18a,wyper.et18b,wyper.et19,doyle.et19} supports the basic concept of
the minifilament eruption model. These workers, however, prefer the term ``breakout model," emphasizing
the external reconnection, which they argue is essential to triggering the eruption.

The situation for active region jets viz {\`a} viz EMFs 
is less certain.  EMFs do make
some active region jets \citep[e.g.,][]{joshi.schmider.et17,solanki.et20}.
There are also studies of active region jets that do not report or observe 
concurrent EMFs, with examples including \citet{cheung.et15} and \citet{mulay.et16,mulay.et18}, 
although it is unclear whether those investigators were specifically looking for such features. 

\citet{sterling.et16b,sterling.et17} also investgated active region jets, and looked
to see whether they were consistent with the minifilament-eruption model.  \citet{sterling.et16b} looked
at several jets from a single active region.  Each of three of those jets that they examined 
in detail displayed a clear EMF at the  start of the jet, erupting from a magnetic neutral line between 
opposite-polarity patches of strong flux.  These neutral lines were either canceling, or undergoing both
emergence and cancelation concurrently.  \citet{sterling.et16b} also,  however, reported observing  several
other jets with cool-material outflows, consistent with a minifilament eruption from a neutral line, but for
which they did not see an actual, well-formed erupting minifilament.

\citet{sterling.et17} looked at coronal jets from a different active region. Again  they found clear EMFs for
several jets. But other jets in the active region, in  particular those from a specific strong-patch subregion
of that active region  (called region ``R2" in that paper) showed copious jets with cool ejected material, but
where a clear, well-formed EMF, could not be detected.  All of the jets of \citet{sterling.et17}, both those
with and those without clear EMFs, occurred along neutral lines undergoing flux cancelation. 
\citet{sterling.et17} investigated closely several  of those jets without clear EMFs, and argued that the
minifilament eruption model magnetic setup  and dynamics for such jets were the same as in the cases where a
clear EMF could be identified (including many non-active-region jets).  The two studies
\citet{sterling.et16b,sterling.et17}, therefore, suggest that the active region coronal jets showing
cool-material outflows but lacking a clear, well-formed EMF likely form via the minifilament-eruption model,
but for some not-yet-understood  reason, a well-developed cool-material minifilament did not form in those
cases.

Nonetheless, for the jets in which we do not
observe an EMF, we cannot rule out the 
possibility that a different
mechanism might operate in some cases, such as the \al ic magnetic twist-wave process without a 
cool minifilament
\citep[e.g.,][]{shibata.et86,pariat.et09,kumar.et18}.  In addtion, in cases where
flux emergence is observed, we cannot rule out that jets may be formed due to the flux-emergence model 
\citep{shibata.et92,yokoyama.et95}.

\citet{moore.et10} and \citet{moore.et13} found that, when observed in soft X-rays, the spires of
jets, roughly  speaking, fall into two categories: in one case, the X-ray spires remain narrow
(compared to the width of the jet's base) throughout life of the jet; they called these ``standard
jets."  In the second case, the X-ray spire broadens during the life of the jet, so that its width
becomes comparable to the width of the base; they called these ``blowout jets."  In this
paper we will again look at a set of polar coronal hole jets in both X-rays and in AIA EUV images.
As with the jets of \citet{sterling.et15}, the jets of this study have EMFs in at 
least 20 of 21 jets; most of these, 11, are blowout jets, while 
four are standard jets, and six are uncertain.

We use these jets to address new points that previously have only been suggested or not 
considered at all
by previous coronal jet studies:  (1) We study the behavior of the EMFs in minifilament eruptions 
that make blowout jets (i.e., jets
whose spires become broad when observed in soft X-rays).  It had been 
suggested previously that blowout jets have EMFs that are ejected; that is, have minifilaments that
erupt out of the jet-base region and escape along the jet spire into the far corona, or even out into the heliosphere.  We will find that this holds for several of the blowout jets of this data set, but 
we will also learn of a new, modified, alternative  behavior for the EMFs in 
some wide-spire jets that has
not been pointed out before, resulting in a modified schematic  picture for some wide-spire jets. 
(2) We examine the behavior of the EMFs that
result in the four standard jets of this study.  Previously, \citet{sterling.et15} 
suggested that standard 
jets (i.e., jets
whose spires remain narrow when observed in soft X-rays) have
mainly confined EMFs, where the EMFs remain largely trapped in the jet-base region.  (\citeauthor{sterling.et15}~\citeyear{sterling.et15} also alternatively suggested that the narrow spires might 
result from eruptions that are ``ejective but very weak," where ``weak" would mean that the EMF's eruption
is very slow and/or faint.)  We will confirm 
that in our four standard-jet cases the EMFs are mostly confined.  (3) We will
highlight that the visibility of the EMFs in jets can vary widely in different AIA channels.  They
can be essentially  invisible in some channels, while plainly visible in others.  This demonstrates that it
is necessary to examine multiple AIA channels before concluding that an EMF might not occur in a
particular jet.  

Before getting into our new data set, we first clarify some points regarding the
history and the meaning of the terms ``standard" and ``blowout" jet.

\section{Standard and Blowout Jets}
\label{sec-sb_jets}


We begin with a brief explanation for the origin of the terms ``standard" and ``blowout" in 
describing jets.

During the time when the emerging flux mechanism was still thought to be the cause of jets, and before
the concept of the minifilament eruption model for jets was developed, \citet{moore.et10} and
\citet{moore.et13} attempted to explain particular characteristics of jets based on the flux emergence
model, and variants of that model.  The characteristics in question were based on  the appearance of the
jets when observed in soft X-rays from XRT\@.  The jets of their studies were restricted to polar coronal
hole locations.  In addition to the X-ray images, the two studies also used \heii\ 304\,\AA\ images, from,
respectively, the EUVI instruments on the \stereo\ satellites  in \citet{moore.et10}, and from AIA in
\citet{moore.et13}.  

In the identification of the two types of jets from the XRT images, \citet{moore.et10,moore.et13} found 
that in the narrow-spire (standard) jets the base brightening generally remains localized to 
the JBP at one side of the jet's base.
In the second type of jet (blowout), \citet{moore.et10,moore.et13} found that the spire starts 
off narrow, but then grows over much of the life of the jet so that the width of the spire becomes comparable to the size of the base.  For these blowout jets, usually the base brightening stops being 
localized to one part of the base, but
instead the entire base evolves to glow with a brightness rivaling that of the JBP\@.  In the 
two \citet{moore.et10,moore.et13} studies, approximately one-half of the jets fell into each of the two categories.

\citet{moore.et10,moore.et13} adopted the ``standard" and ``blowout" names based on the idea that both  jet
types were a  consequence of the emerging flux model.  In the case of the narrow-spire jets, they assumed
that  the jets were due to the process depicted in drawings in \citet{shibata.et92}.  The emerging-flux
bipole would reconnect with surrounding open coronal field \citep[this would be external reconnection, in
the language of][]{sterling.et15}.  This external reconnection was argued to reconnect the 
emerging-loop field with open field lines, with some of
the closed emerging-loop field lines becoming reconnected to open field 
lines by the external reconnection.  This
reconnection also would both heat the coronal plasma, and eject that heated material vertically along the
reconnected open field; this would result in the X-ray-jet spire, with the spire emanating from the  location of the
external reconnection and flowing out along a narrow column.  According to this emerging-flux picture, a
second product from that external reconnection would be in the form of a new closed loop below the reconnection
site; it was envisioned that this would be the JBP brightening at an edge of the base of the jet. In this case, it was
imagined that the emerging bipole remained in tact, making narrow-spire jets when enough oppositely directed
field between the emerging bipole and the ambient corona built up for reconnection to ensue.
\citet{moore.et10,moore.et13} envisioned that the narrow-spire jets followed this scenario, and therefore
they named these jets standard jets, simply because they envisioned that they followed the standard scenario
proposed by \citet{shibata.et92}.   

For the broad-spire jets, \citet{moore.et10,moore.et13} envisioned that initially the evolution
is the same as in the standard-jet case.  But in this case, before, at, or after the initial external
reconnections,  the emerging bipole itself becomes unstable and erupts outward into the overlying
corona.  This exploding bipole would erupt outward, and material heated via more widespread
external reconnections during the eruption  would become the jet's spire.  Because the
erupting bipole would have an extent comparable  to the extent of the pre-eruption bipole that
forms the base of the jet, this scenario would  explain the broad width of the jet spire in
these cases.  Moreover, internal reconnections of the legs of the erupting emerged bipole would lead
to heating and brightening of the entire base region, explaining the increased extent of the
base brightening in these cases.  (Again after
\citeauthor{sterling.et15}~\citeyear{sterling.et15}, we call these reconnections {\it internal}
reconnections, because they occur internal to the closed bipole.)  Because the bipole erupts
open in this case, \citet{moore.et10,moore.et13} called these blowout jets.

Thus, the original concept of ``standard jet" and ``blowout jet" was based on the idea that jets basically
resulted according to the  emerging-flux idea.  In a variant of the original emerging flux model,
\citet{moore.et10,moore.et13} also allow for the possibility that the bipole had stopped emerging, but
that it still created the jet when the emerged bipole undulated or was otherwise
disturbed, so that the core of the bipole starts to erupt and/or a burst of external 
reconnection begins between the bipole and the surrounding ambient coronal field. 
Thus, in the original idea of \citet{moore.et10,moore.et13}, both standard and blowout jets were a
consequence of either an emerging, or an already emerged, magnetic bipole undergoing a sudden
episode of fast external reconnection
with surrounding coronal field; and the JBP was a consequence of this external reconnection,
occurring below the site of the external reconnection.  

All of this explanation from \citet{moore.et10,moore.et13} for the cause of standard and blowout jets,
however, was before  we realized that many (if not all) jets result from minifilament eruptions from  the site
of the JBP\@.  Therefore, we had to modify our suggestion of what made the narrow-spire and the wide-spire
jets.  In \citet{sterling.et15}, we therefore argued that both the narrow-spire jets and the wide-spire jets
are made by minifilament eruptions.  \citet{sterling.et15} further suggested that the difference came about
based on whether the EMF was confined to the base, in which case the spire would be narrow (and hence a
``standard jet"), or ejected so that it explosively escaped from the base (making a broad-spire ``blowout
jet").  Thus, the term ``standard jet" should be considered a misnomer: that name implies that the
narrow-spire jets are formed by the emerging-flux model, which \citet{moore.et10,moore.et13} had regarded as the
standard model for jets.  But  the evidence supports that those standard jets (and blowout jets too) result
from minifilament  eruptions, at least for  many jets.  (Perhaps ``narrow-spire jet" and ``broad-spire jet"
would have been more descriptive names for these types of jets, but the names ``standard jet" and ``blowout
jet" are now historically  tied to these respective jet types.)

So in summary: the evidence we have currently is that both X-ray broad-spire jets 
(named blowout jets), and X-ray narrow-spire jets (named standard jets) result from minifilament
eruptions.

One of the topics of our current study is to examine the behaviors of the EMFs in the blowout jets
(i.e., our X-ray-observed broad-spire jets) and in standard jets (i.e., our X-ray-observed narrow-spire jets)
in our data set. For this set of events, we will examine in closer detail than before the behavior of the
EMFs in these jets.  Specifically, we will ask whether our blowout jets  always have ejective
EMFs, and whether the four standard jets  of our study have confined EMFs.  This is not meant to be
comprehensive statistical study of this question, but, using the jets of each type in our set, for 
the first time we will study the behavior of the EMFs in the jets.

\section{Instrumentation and Data}
\label{sec-data}

We use data from \hinode/XRT and \sdo/AIA for this study.  \hinode\ \citep{kosugi.et07} 
was launched in 2006, and observes the Sun from Earth orbit.  XRT \citep{golub.et07} observes
the Sun in soft  X-rays, over a range spanning a few~\AA\ to $\sim$100~\AA, using a variety of
filters.  For this study we restrict our observations to those taken with the ``Al Poly"
filter, as  it is sensitive to relatively cool temperatures in the range of  $\sim$10$^6$\,K
\citep{narukage.et14}.  These images are appropriate for observing X-ray jets in polar
coronal holes, which often have spire temperatures in the  $\sim$1---2\,MK range
\citep{pucci.et13,paraschiv.et15}.   XRT is capable of observing the Sun with various
parameters.   For our observations, XRT typically used 60\,s or slightly faster cadence, 
with $1''$ pixels and a field of view (FOV) of $384'' \times 384''$.


\sdo/AIA  observes the Sun with seven EUV filters, identified by the
wavelengths 304, 171, 193, 211, 131, 94, and 335\,\AA, listed in roughly
ascending order from lower to higher temperature sensitivity for non-flaring plasmas,
covering $\sim$10$^5$---10$^7$\,K \citep{lemen.et12}.  \citet{sterling.et15} 
concluded that the hottest three channels: 131, 94, and 335\,\AA, added little to 
the investigation of the processes driving the jets observed in polar coronal holes.  
Therefore for this study we only include the first four channels in our data.  AIA 
observes the entire solar disk with 12\,s cadence in these filters, with pixels of 0.$''$6.

Our observed jets can sometimes have weak signal in our selected passbands.  For
the work here we have increased the signal by summing two consecutive 
images for both our XRT and AIA images.  We formed the presented accompanying 
videos as running sums of the images, maintaining the 12\,s cadence for the
AIA movies and about 1\,min cadence for the XRT movies.  Additionally, we have
enhanced visibility of faint features by taking the fourth root of the intensities,
and we present the resulting images and videos with a typical logarithmic scaling.

We selected the jets for our study from the XRT images rather than from AIA\@.    The
reason for this is two fold. First: We are interested specifically in observing 
X-ray jets, because that is what was used to identify standard and blowout jets in
the \citet{moore.et10,moore.et13} studies.  But jets  that stand out in EUV
wavelengths may not necessary be prominent, or even visible  at all, in X-rays. Thus,
to be certain that we are observing X-ray jets, it is necessary to start with the
X-ray images.  Second: Because the XRT data are not of the full-disk  synoptic form
of AIA, first selecting jets with AIA would not guarantee that XRT was observing the
region of interest at the same time.  But AIA data are almost certainly available for
any of the selected XRT time periods, provided that the data are from after AIA
started operations in 2010; thus any of the jets we select from XRT images after that time 
will almost certainly have corresponding AIA images.

We selected time periods when \hinode\ was running the \hinode\ {\it operations
program} (HOP)~81, which  has been regularly done by the \hinode\ team beginning early in
the mission.  This HOP observes
the polar coronal hole regions, with XRT typically taking observations in
selected filters for several consecutive hours.  We looked for data during 2017
and 2018, when the polar coronal holes were present again following the maximum 
of solar cycle~24.  For ease of analysis, we
searched for periods of continuous observations over several hours in which it
was obvious from low-resolution quick-look data that several jets would be
included.  Our aim was not to do a completely random statistical study of  all
jets, but rather we were looking for a sizable number of clear X-ray jets to
investigate together with AIA data.  We settled on data sets from four 
dates: 2017 January~13, 2017 March~6,
2017 July~4, and 2018 April~3. Our resulting data set includes 21
jets in polar coronal holes that were well observed by XRT; as expected,
AIA data are available for all of these events.  Due to the near-pole location
of these jets, we did not attempt to analyze corresponding magnetic field 
data, as the signal for such near-limb non-active-region locations is usually 
too weak to be useful.

\begin{figure}
\hspace*{0.0cm}\includegraphics[angle=0,scale=0.5]{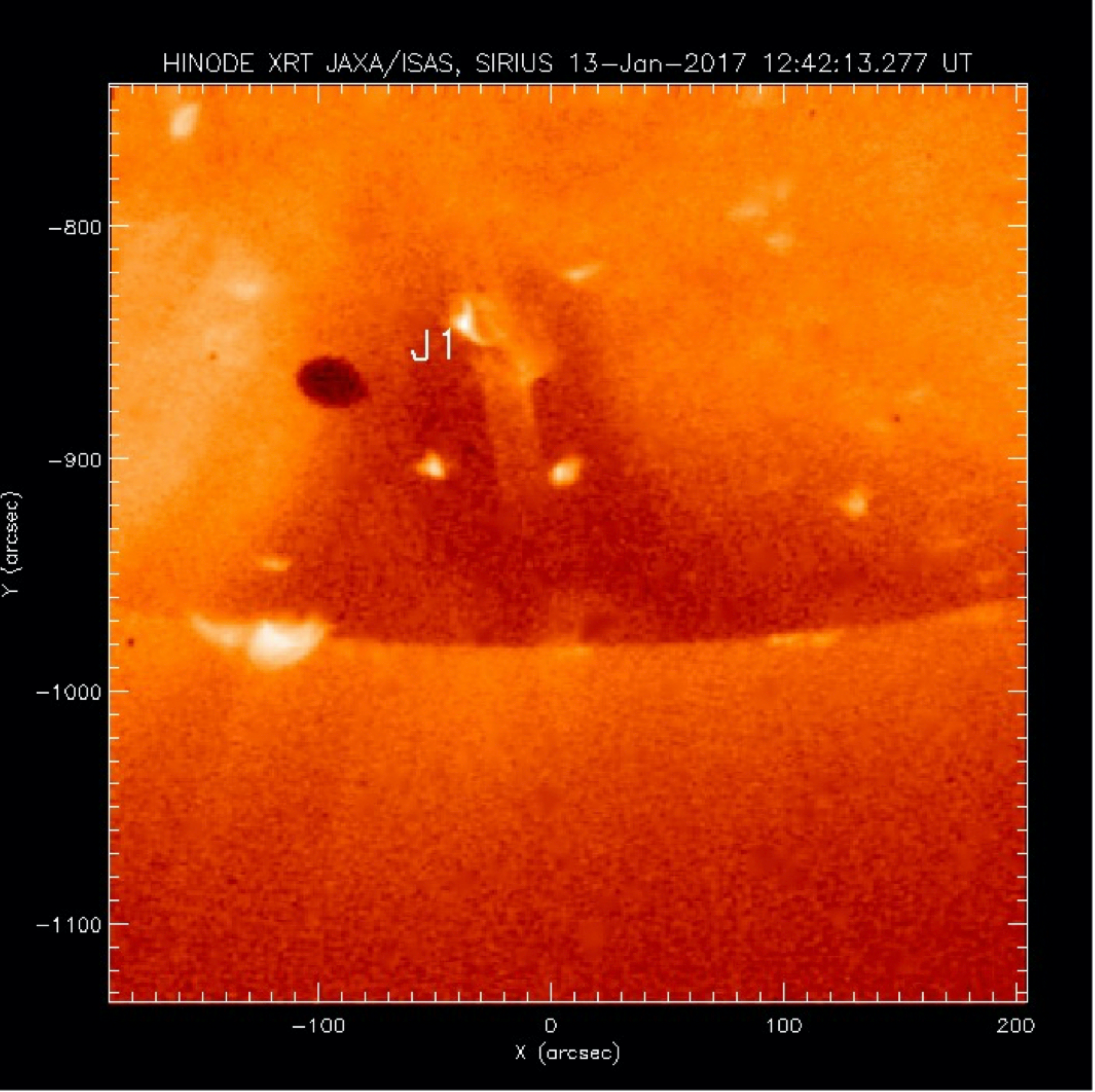}
\caption{
A soft X-ray image from the \hinode\ XRT instrument's Al~Poly filter, of the south polar coronal 
hole region on 2017 January~13.  One of the jets of
Table~1, labeled J1, is in progress.   North is up and west is to the right in this and in all 
other solar images and animations of this paper.  The dark oval at about (-80,-870), and similar 
smaller patches and spots, are artifacts. The accompanying animation shows the XRT movie for
this period, covering 2017 January 13 over
12:00:36---13:58:02\,UT\@.  It is constructed by performing a running sum of every
two consecutive images, and is presented as a full-cadence (60\,s) running-sum movie. The movie
plays twice, first showing Jets J1---J4 labeled as they happen, and then again without labels. The 
entire movie runs for 12\,s.
}
\label{f_xrt1}
\end{figure}

\begin{figure}
\hspace*{0.0cm}\includegraphics[angle=0,scale=0.5]{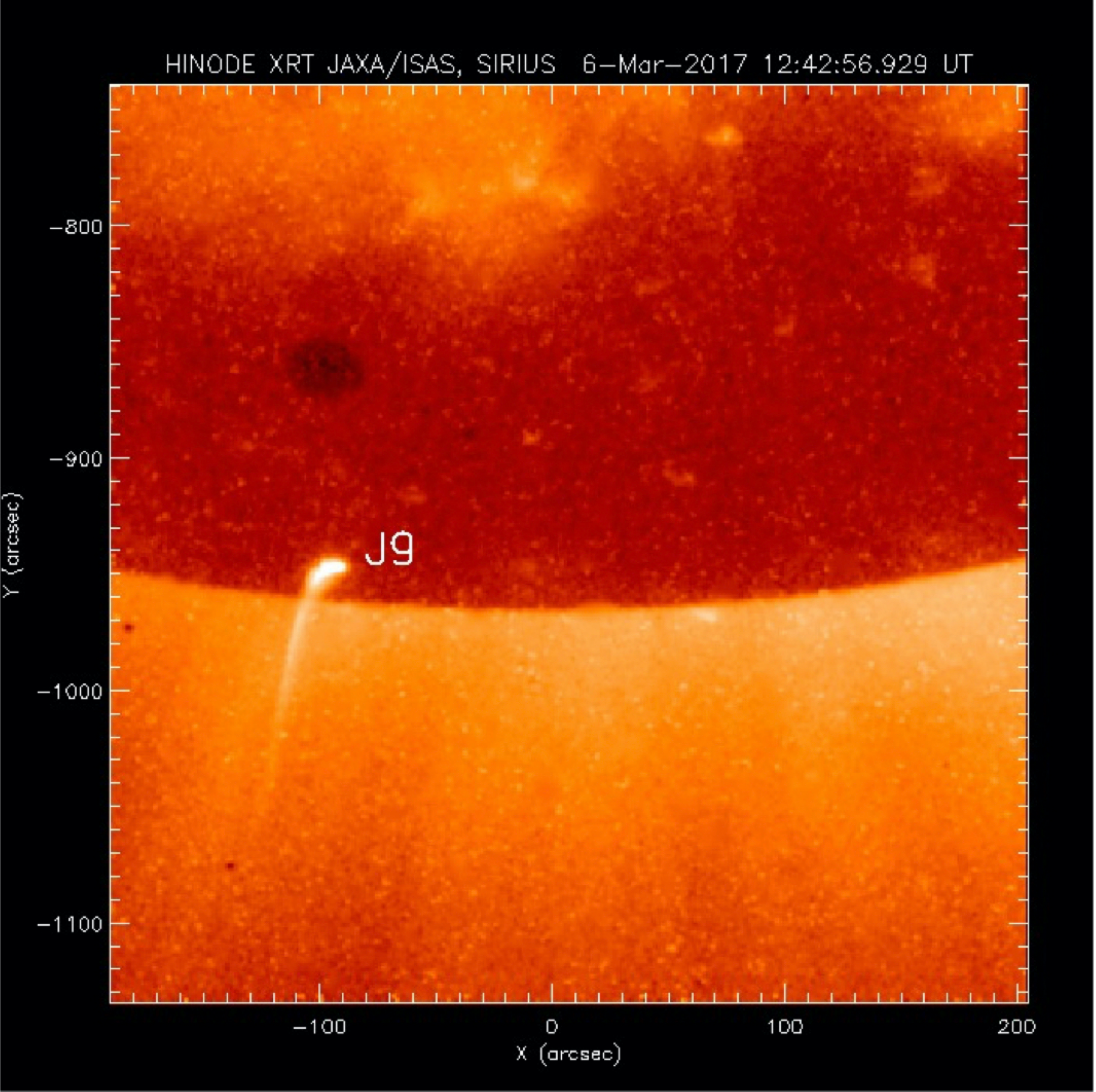}
\caption{
Similar to Fig.\,\ref{f_xrt1}, but showing the south polar coronal hole region 
on 2017 March 6, at a time when jet J9 of Table~1 is in progress.  The 
dark oval at about (-80,-870), and similar smaller patches and spots, are artifacts.
The animation covers 2017 March 6, over
10:34:30---15:58:06\,UT\@. It is constructed by performing a running sum of every
two consecutive images, and is presented as a full-cadence (60\,s) running-sum 
movie. The movie plays twice, first showing Jets J5---J14 labeled as they occur, 
and then again without labels. The entire movie runs for $\sim$31\,s.
}
\label{f_xrt2}
\end{figure}

\begin{figure}
\hspace*{0.0cm}\includegraphics[angle=0,scale=0.5]{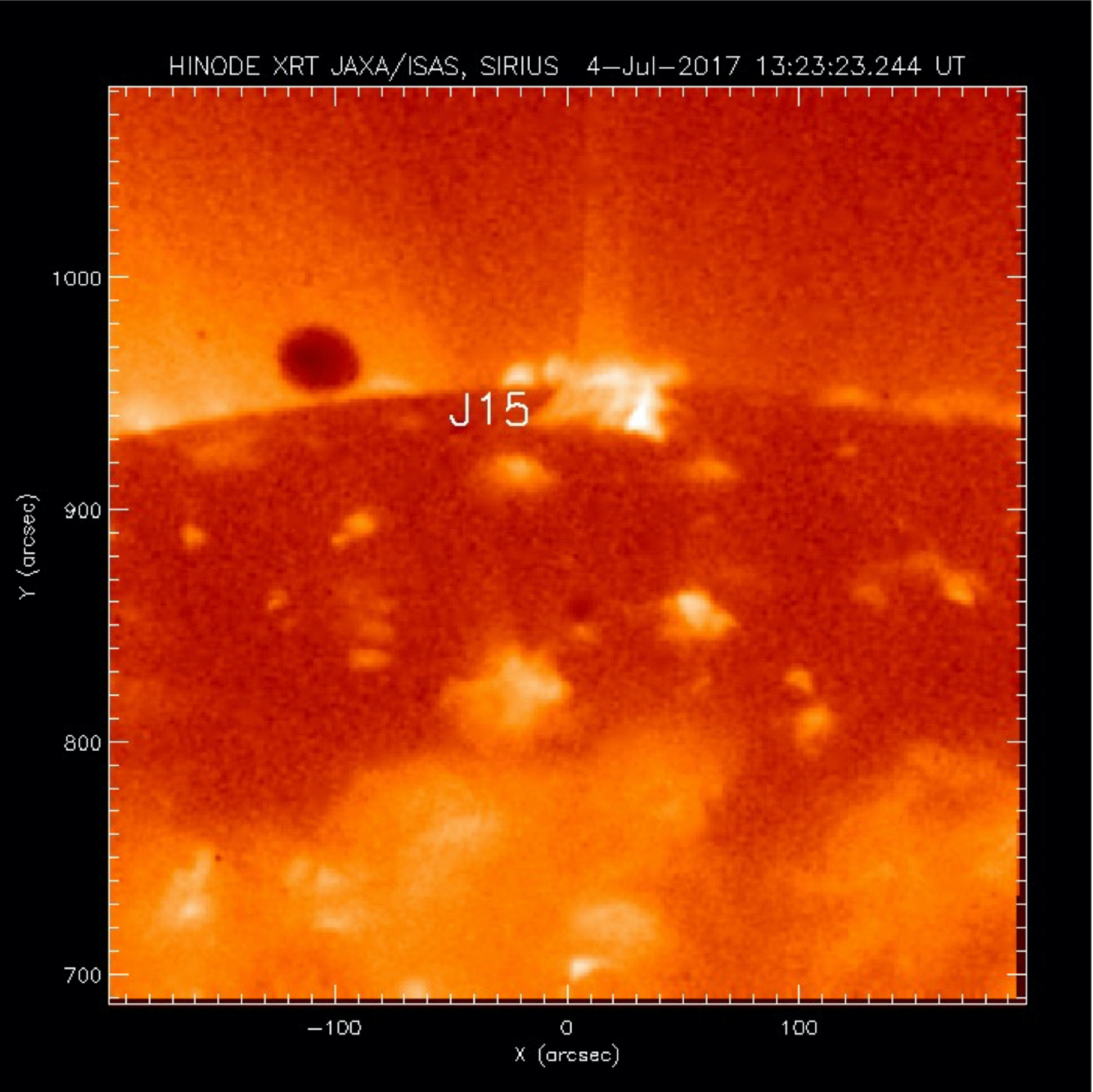}
\caption{
Similar to Figs.~\ref{f_xrt1} and~\ref{f_xrt2}, but showing the north polar coronal hole region on 2017 July 4, at a time when jet J15 of Table~1 is in progress.  The dark oval at about (-100,970), 
and similar smaller patches and spots, are artifacts. The animation covers 2017 July 4 over
12:45:05---16:33:20\,UT\@. It is constructed by performing a running sum of every
two consecutive images, and is presented as a full-cadence (60\,s) running-sum movie. The movie
plays twice, first showing Jets J15 and J16 labeled as they occur, and then again without labels. The entire
movie runs for $\sim$29\,s.}
\label{f_xrt3}
\end{figure}

\begin{figure}
\hspace*{0.0cm}\includegraphics[angle=0,scale=0.5]{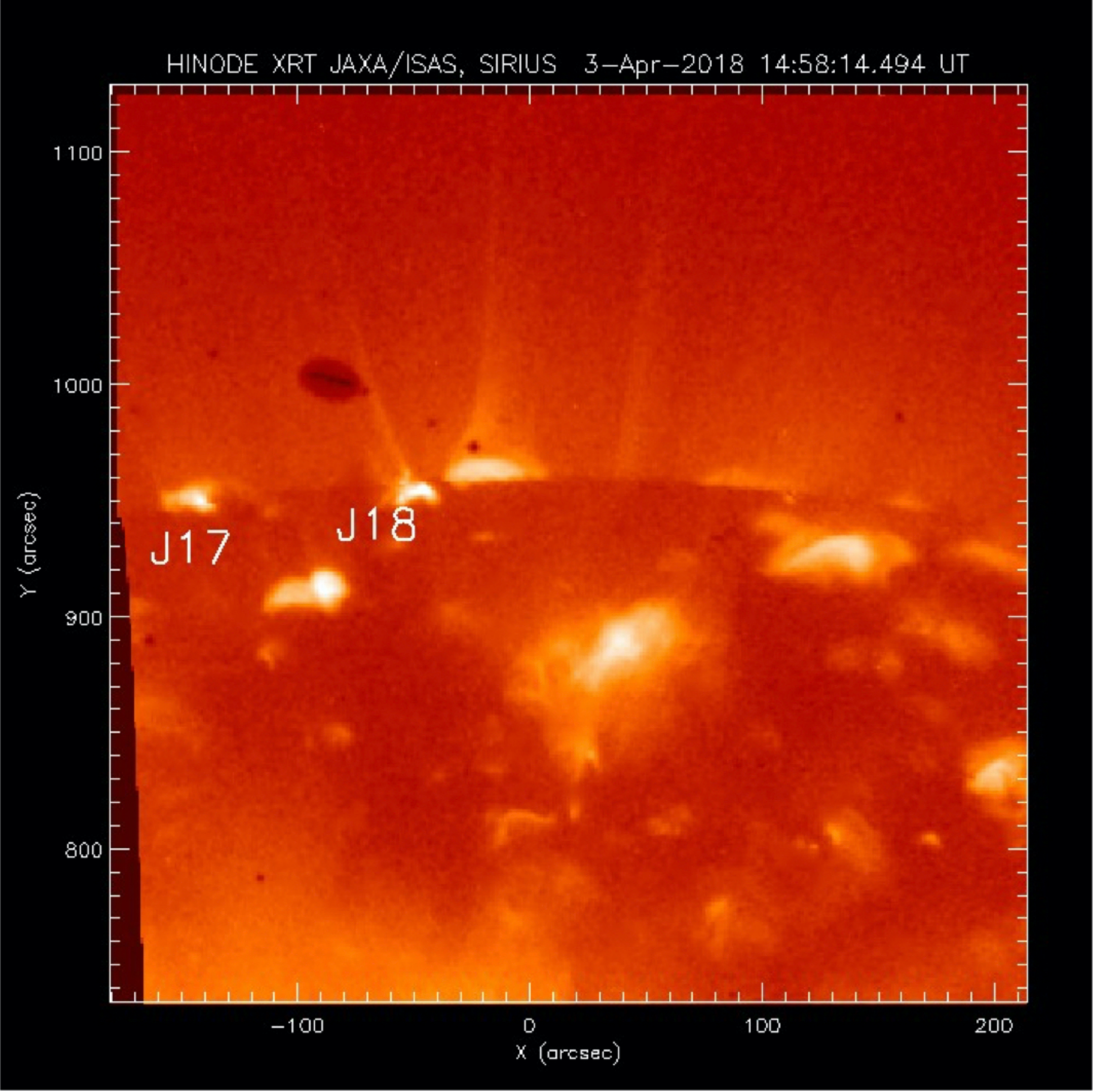}
\caption{
Similar to Figs.~\ref{f_xrt1}, \ref{f_xrt2}, and~\ref{f_xrt3}, but showing the north polar coronal hole region on 2018 April 3, at a time when Table 1 jet J17 is about to end, and jet J18 is in progress.  Also apparent in this image is a jet at the limb near $x=0$; although prominent in X-rays, we feared that it is too close to the limb for a reliable analysis of the corresponding AIA images, and therefore we did not include it in our study.  The dark oval at about (-100,970), and similar smaller patches and spots, are artifacts.  The animation covers 2018 April 3, over
13:48:58---17:36:02\,UT\@.  This video is constructed by performing a running sum of every
two consecutive images, and is presented as a full-cadence (60\,s) running-sum movie. The movie
plays twice, first showing Jets J17 and J21 labeled as they occur, and then again without labels. 
The entire movie runs for $\sim$30\,s.}
\label{f_xrt4}
\end{figure}

Table~1 lists our selected jets, which we have labeled chronologically as J1---J21. 
Figures~\ref{f_xrt1}, \ref{f_xrt2}, \ref{f_xrt3}, and~\ref{f_xrt4} show XRT 
snapshots for the four time periods, with some of the jets labeled.  Accompanying
animations show the complete periods of observation, with all of the jets labeled. 
As can be ascertained from the videos, we selected only a selection of all the jets
visible in each animation.  We selected most of the obvious strong jets, but for
various reasons omitted some jets that were prominent in XRT\@.  One  such reason is
that some jets very near the limb did not seem appropriate for analysis  after a
cursory inspection of the corresponding AIA images.  For example,  the jet in
Figure~\ref{f_xrt4} west of Jet J18, near coordinates (0,950), is prominent in
XRT\@.  When we looked at the corresponding AIA images however, we were not confident
that we were able to observe adequately the near-photosphere dynamic activity 
leading to the
jet, and therefore we did not include that jet in our study. Another reason for why
we omitted some otherwise-attractive XRT jets was because of a gap in the  continuity
of one of the data sets.  While most of the selected jets stand out in XRT images,
for variety we have included a few jets that are less  prominent yet still distinct
in XRT; jets~J7, J8, J12, and~J19 being such jets.  

Columns two and three of Table~1 give the date and time of the start of each jet, and
column four gives the jet duration.   One has to make a choice in deciding when a jet
starts.  We have  opted to use as the start time the XRT frame in the videos
accompanying Figures~\ref{f_xrt1}---\ref{f_xrt4} that first shows the spire of the jet
starting to grow.  This too can be somewhat subjective, and thus there is an
uncertainty in the start time of approximately one frame of the XRT movies. 
Similarly, we define the duration based on when the spire is no longer visible. 
Another equally acceptable option for defining duration would have been to  use the
time that the  base starts to brighten  as the jet start time; doing so would
increase the estimated duration over that obtained via our adopted method by several
minutes in some cases.  

Column~4 of Table~1 gives our assessment of each jet as standard, blowout, or uncertain, 
based on
the width of its spire in the XRT images.  We will discuss these aspects of the jets
in \S\ref{sec-s_and_b}.  Before that however we will consider the activity at the
base of the jets in \S\S\ref{sec-emf} and \ref{sec-characteristics}.

\section{Erupting Minifilament Behavior at Jet Bases}
\label{sec-emf}

\subsection{Over-and-Out Erupting Minifilaments}
\label{sec-oao}

We have found that the appearance of the EMF is different in different jets.  One variety 
are what can be called ``over-and-out" EMFs. They basically fit the picture of 
\citet{sterling.et15}, where the
minifilament starts at  one side of  the large lobe, and migrates over to near the
top of that lobe, and then largely  escapes through newly opened field there.  These
eruptions are similar to a larger class of over-and-out eruptions, ranging from jet features to
CMEs \citep[e.g.][]{bemporad.et05,moore.et07,sterling.et11b,gopal.et14,panesar.et16b,yang.et18}.

\begin{figure}
\hspace*{0.0cm}\includegraphics[angle=0,scale=0.73]{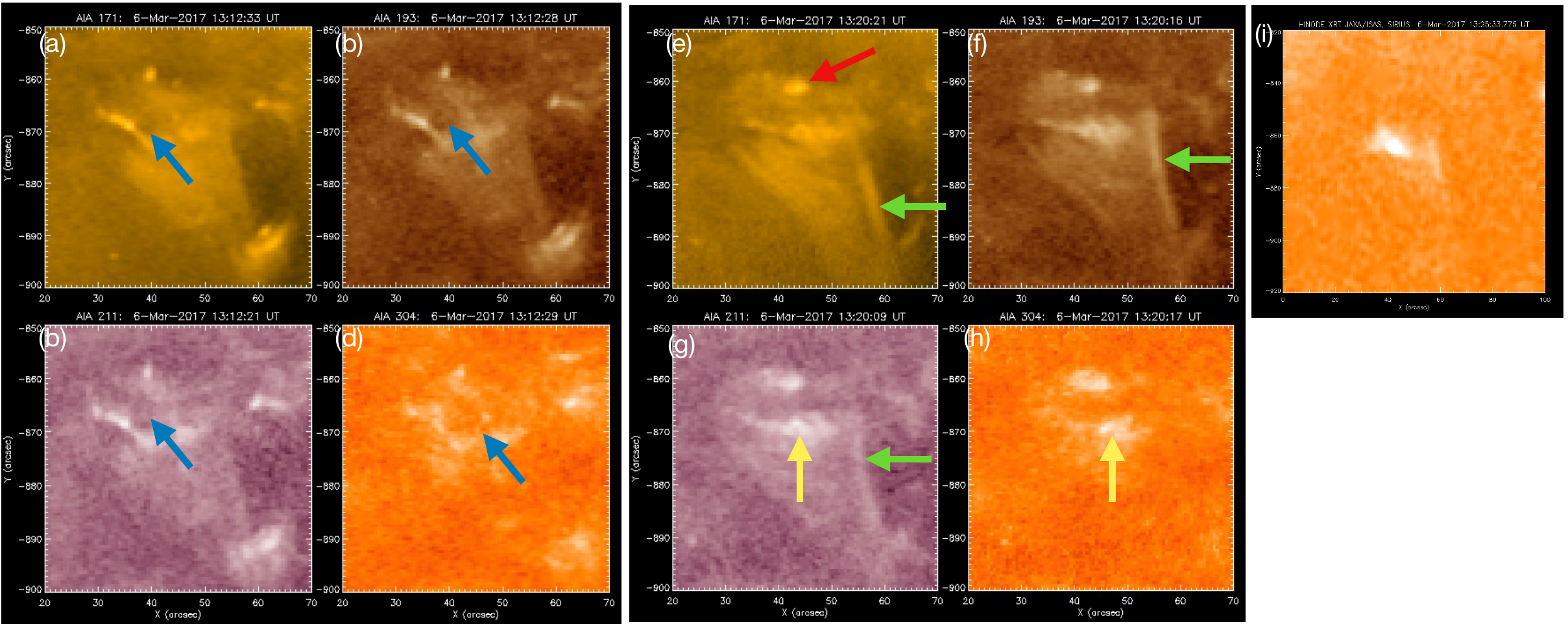}
\caption{
AIA images and an XRT closeup of Table~1 jet~J11.  The AIA images are in 171, 193, 211, and 304~\AA, 
showing the jet at two different times,
near 13:12~UT (left four panels) and near 13:20~UT (right four panels, on 2017 March~6; details of each
frame are provided in the labels at the top of the panels.
Blue arrows show an erupting minifilament (EMF) in absorption.  By the time of the right-side
panels the minifilament has erupted out of the field of view.  The red arrow shows the location
of the JBP, corresponding to the location between M2 and M3 in Fig.~1 of \citet{sterling.et18} (and in Fig.\,\ref{f_schematic2} below).  
The yellow arrows show the location of the large lobe, corresponding to the location between M1 and M2 in those schematics.  It has brightened due to the EMF flux rope's
external reconnection.  In the left four panels, this lobe is partially obscured by the erupting cool minifilament in the foreground.  Green arrows point to the jet spire. Panel~(i) shows a zoomed-in XRT
view of the jet (axis ranges are different from those of the AIA panels). The animation covers
2017 March 6, over approximately 12:50---13:50\,UT for AIA, and over 13:08---13:50\,UT for XRT\@. 
The videos are constructed by performing a
running sum of every two consecutive images; the AIA videos are presented as full-cadence (12\,s)
running-sum movies, while the XRT video is presented as a running-sum movie of cadence $\sim$60\,s.  
The entire five-panel movie runs for $\sim$7\,s.
}
\label{f_j11}
\end{figure}

We show two examples from this over-and-out EMF category.  Figure~\ref{f_j11} shows the
first  of these examples, which is event~J11 of Table~1.  This EMF is visible in
all four of our AIA channels.  We display our example jets in four-panel image arrays, where
panels~(a), (b), (c), and (d) respectively show the same FOV in the AIA 171, 193,
211, and 304\,\AA\ channels, at approximately the same time. Panels~(e), (f), (g),
and (h) show the same for a different time period. The corresponding video
shows the same images over the duration of the event.  Panel~(i) and the corresponding
video show the jet in XRT images (with a different FOV from the (a)-(h) AIA
images).  Early movement of the minifilament  (blue arrows in Fig.\,\ref{f_j11})
becomes obvious from about 13:07~UT (see the corresponding video).   The feature  travels
toward the southwest from then until about 13:18~UT,  when its trajectory changes to
a more southerly direction, and then travels in that direction  outward along the
jet's spire (green arrows).  In this case, pre-jet ambient open field lines along the
line of sight to the newly reconnected open field lines upon which the spire forms apparently
already contain material prior to the event, so that they form a plume-like structure
visible in the  the 171, 193, and 211~\AA\ channels in the (a)---(d) panels.  An
enhancement flows out along the western edge of the structure at the time of the
right-side panels  in the figure and accompanying video, forming the spire.  In XRT
(animations in Figs.~\ref{f_xrt2} and \ref{f_j11}), this jet is comparatively weak; its JBP is
likely faintly visible in the frames at 13:15 and 13:16~UT, although it is more
apparent in  the AIA images (red arrow in Fig.\,\ref{f_j11} points to it). From those
AIA images, and also from the XRT videos as the spire becomes more visible, at about
13:23~UT for example, it is apparent that the JBP occurs on the northeast  side of
the spire.  This is consistent with the JBP forming at the location from where the
minifilament erupted in the AIA images, consistent with 
\citet{sterling.et15}.  The brightening of the large lobe from the external
reconnection  (straddling M1---M2 in \citeauthor{sterling.et18}~\citeyear{sterling.et18}; also see Fig.\,\ref{f_schematic2}, below) is apparent in XRT
images from about 13:18~UT, and from about 13:20~UT in the AIA images (yellow
arrows).  In this case, this lobe outshines the JBP, consistent with a common
occurrence in blowout jets reported by \citet{moore.et10,moore.et13}.

\begin{figure}
\hspace*{0.0cm}\includegraphics[angle=0,scale=0.7]{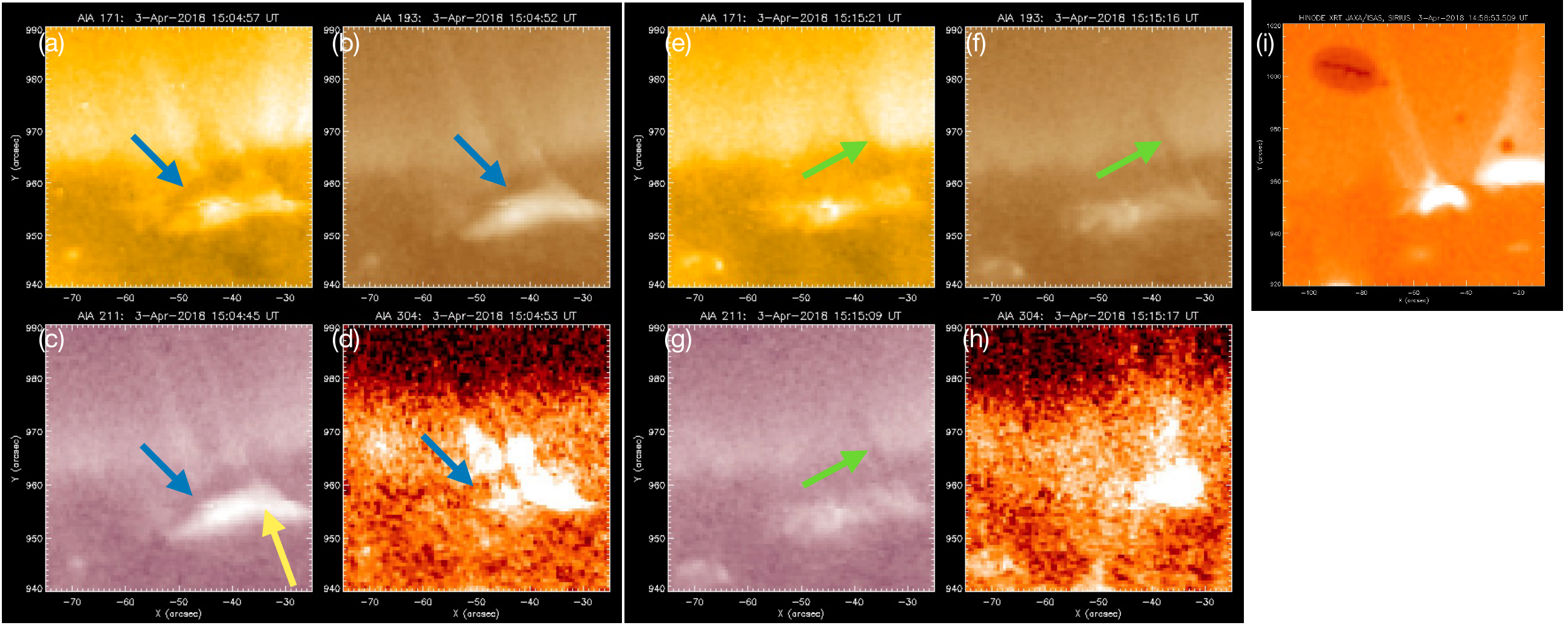}
\caption{
AIA images and an XRT closeup of Table~1 jet~J18. The AIA images are in 171, 193, 211, and 304~\AA, showing the jet at two different times,
near 15:04~UT (left four panels) and near 15:15~UT (right four panels, on 2018 April~3; details of each
frame are provided in the labels at the top of the panels.  Arrow colors are as in Fig.\,\ref{f_j11}. From the accompanying animation, the minifilament initially
resides on the far side of the big lobe, and at the time of panels (a)---(d) it has crawled up to
the top of that lobe. In panels (e)---(h), the minifilament is erupting outward; in this case, the green arrows show a strand of dark minifilament material in the spire, while brighter strands of the 
spire are apparent in the XRT movies (Figs.~\ref{f_xrt4} and \ref{f_j18}).  Panel~(i) shows a zoomed-in XRT
view of the jet (axis ranges are different from those of the AIA panels);
the large dark oval, and similar smaller patches and spots, are artifacts. 
The animation covers
2018 April 3, over approximately 14:34---15:20\,UT\@. 
The videos are constructed by performing a
running sum of every two consecutive images; the AIA videos are presented as full-cadence (12\,s)
running-sum movies, while the XRT video is presented as a running-sum movie of cadence $\sim$30---60\,s.  
The entire five-panel movie runs for $\sim$7\,s.}
\label{f_j18}
\end{figure}

In Figure~\ref{f_j18}, Jet~J18 shows another example of this type of jet-producing
minifilament eruption, this time viewing the phenomenon in profile (side-on).  In this 
case, our perspective is looking from the left side of the schematic of 
\citet{sterling.et15}, from a low angle above the
photospheric level.  From the animation accompanying the figure, 
a minifilament (or at least one component of a minifilament) initially resides on the 
far side of the large lobe (the lobe between M1 and M2 in Fig.~\ref{f_schematic2}).  
As it starts erupting, this minifilament closely hugs the
curve of the M1---M2 lobe, as in the four left-hand panels of Figure~\ref{f_j18}. 
That video also shows a JBP brightening at about 14:55~UT\@.  Some minutes  later in
the right-hand panel, at about 15:15~UT, much of this minifilament is erupting  
outward.  Green arrows in Figure~\ref{f_j18}
show the cool-component strands of that EMF\@.  Prior to that time,
from about 14:58~UT until at least 15:08~UT, there are brightenings visible in the
171, 193, and 211~\AA\ channels  corresponding to the hot spire, {\it sans} the cool
material; in the corresponding XRT movie, a bright X-ray spire is visible
over about 14:51---15:08~UT, and these hot-spire observations correspond to the middle
panel (panel b) of the schematic in \citet{sterling.et15}.  Thus,  all of the key aspects of the
\citet{sterling.et15} schematic are apparent in this jet.  (This description however
may be oversimplified somewhat, as another minifilament component, in the foreground of the component of
the minifilament just discussed, is apparently 
undergoing eruption between 14:49 and 14:59~UT\@.  Due to the single viewing perspective, it
is not possible to sort out the detailed connection between these different 
minifilament components.)

There are several other examples of the same basic type of eruption in our data set. 
For example,  jets~J4, J13, and J14 all apparently belong to this category of
over-and-out EMF\@.  Each of these cases show well the minifilament
corresponding to times  between the those represented by the first two panels 
(panels~(a) and~(b)) of the \citet{sterling.et15} schematic. 

\subsection{Largely Confined Erupting Minifilaments}
\label{sec-rolling}

A second type of EMF that we observed in our data set is one
where the in-motion minifilament appears to be confined very low near the
surface, as it travels on top of the magnetic lobe between locations M1 and M2
in Figure~\ref{f_schematic2}(a).  In many cases it can appear to execute a
revolving type of motion as it travels over the top of that lobe in a direction
away from its ejection location.  In our data set,  jets~J5, J16, J18, J20, and
probably J17, exhibit this type of behavior.  Here we present two examples; we
present jet J5 because it shows the phenomenon well, and we present J20 because it
gives key insight into what is occurring.  We are forced to say that these are ``largely
confined," because frequently we cannot confirm that all of the cool minifilament 
material remains fully confined; some of it may ``leak out" along opening field lines,
or in some cases it might be ejected directly toward us (and hence hard to observe)
later in its evolution.  But at least for much of the duration of these minifilament
eruptions, a substantial portion of the cool material appears to be confined near the
solar surface.

\begin{figure}
\hspace*{0.0cm}\includegraphics[angle=0,scale=0.85]{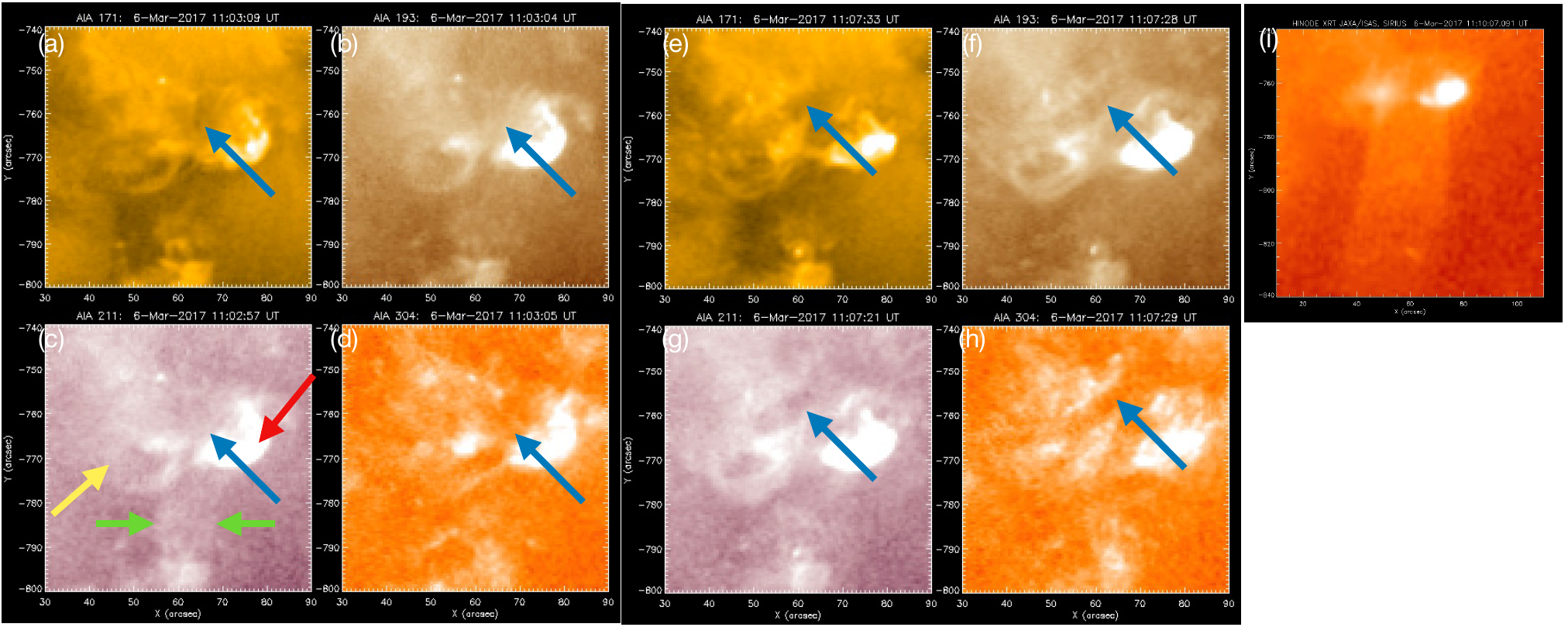}
\caption{
AIA images and an XRT closeup of Table~1 jet~J5.  The layout is as in Figs.~\ref{f_j11} and \ref{f_j18}, but for event~J5
near 11:03~UT (left four panels) and near 11:07~UT (right four panels, on 2017 March~6.  
Arrow colors are as in Fig.~\ref{f_j11}.  From the accompanying animation, 
in this case the minifilament initially
resides on the southwest side of the M1---M2 lobe of Fig.~\ref{f_schematic2} (below), erupts and
rotates over that lobe to its northeast side. Apparently it erupts and largely remains trapped in
the base of the region as depicted in Fig.\,\ref{f_schematic2} (at least until $\sim$11:09~UT, after which
time it we cannot determine with certainty whether it remains confined or ejects outward).  Panel~(i) 
shows a zoomed-in XRT view of the jet (axis ranges are different from those of the AIA panels). 
The animation covers
2017 March 6, over approximately 10:34---11:30\,UT\@. 
The videos are constructed by performing a
running sum of every two consecutive images; the AIA videos are presented as full-cadence (12\,s)
running-sum movies, while the XRT video is presented as a running-sum movie of cadence $\sim$60\,s.  
The entire five-panel movie runs for $\sim$9\,s.}
\label{f_j05}
\end{figure}

Figure~\ref{f_j05} shows the same four AIA channels as in the previous examples, this
time for jet~J5 of Table~1.  As is apparent from the accompanying video, in
this case the minifilament starts erupting at about  11:01~UT from about $y$ location
-770.  An absorbing clump of it is visible passing over the location of
Figure~\ref{f_j05}(a---d) around 11:03~UT, and then seems to settle on the NE side of
the frame near the end of the video at 11:30~UT\@.  Thus, it seems as if the
minifilament revolves around an axis in the plane of the image that runs along a SE-NW
diagonal in the frame.  From 211~\AA\ panel (Fig.\,\ref{f_j05}(c), and also but less
apparent in Fig.\,\ref{f_j05}(d)), the spire of  the jet emanates from that revolving
EMF; this is what causes the broad  spire in the corresponding XRT
video.  This apparently corresponds to the case of the schematic in 
\citet{sterling.et15}, but where the minifilament travels beyond the apex of the 
large lobe field.  The corresponding large lobe is faintly apparent in AIA 211~\AA\
images (yellow arrow in Fig.\,\ref{f_j05}(c)), although it is much more apparent in
the XRT video over about 10:57---11:09~UT\@.  In this case, only a hot spire is
visible. At least up until this time (11:09~UT), little if any of the cool minifilament 
material appears to have been ejected outward in this case.  After  this time it is
hard to discern the minifilament's status; much of it might have been  thrown to the
far side of the large-lobe field, or alternatively much of it may have encountered,
reconnected with, and escaped along open field directed toward the observer (making
that ejection hard to discern), in which case it would, eventually, be of the 
over-and-out variety.  We can say that the X-ray  spire is  prominent
between about 11:00~UT and 11:23~UT, and so its earliest (and strongest expansion)
phase occurred when much of the EMF was still confined to a
comparatively low altitude. 

\begin{figure}
\hspace*{0.0cm}\includegraphics[angle=0,scale=0.65]{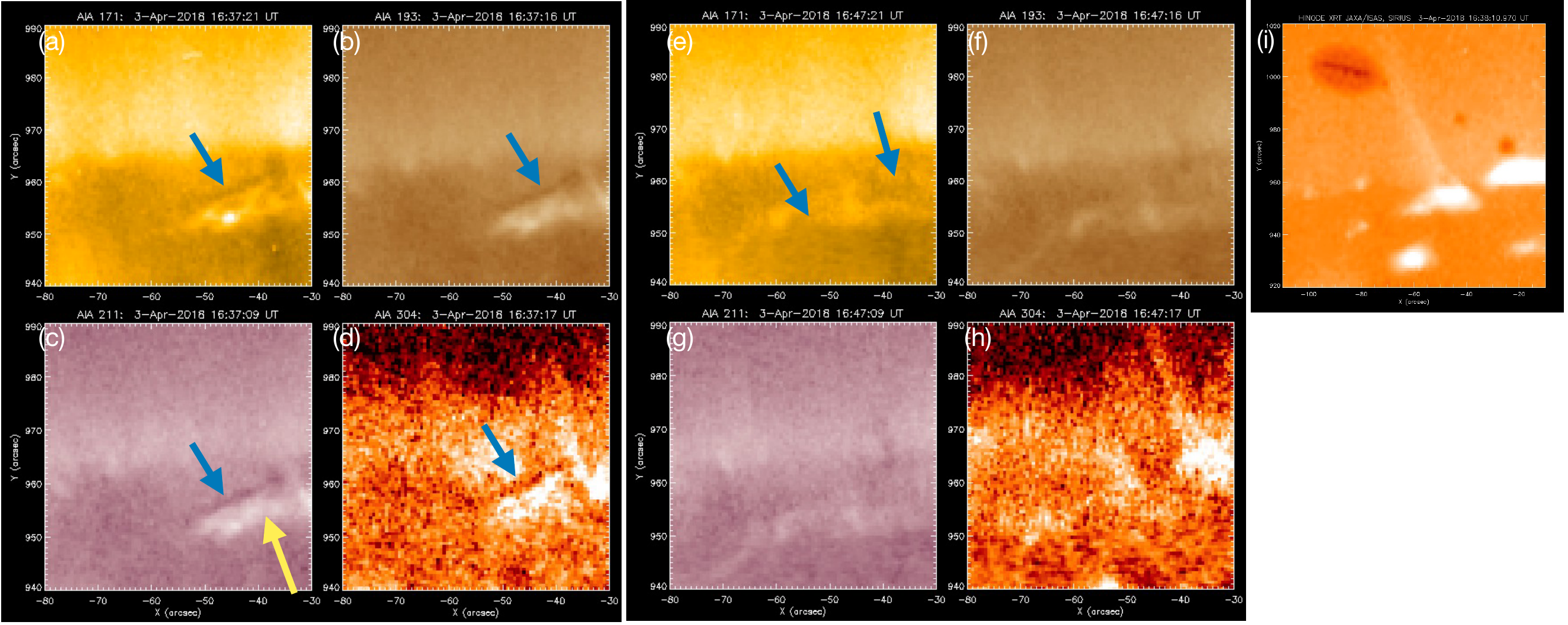}
\caption{
AIA images and an XRT closeup of Table~1 jet~J20.  The layout is as in 
Figs.~\ref{f_j11}---\ref{f_j05}, but for jet event~J20
near 16:37~UT (left four panels) and near 16:47~UT (right four panels, on 2017 March~6.  
Arrow colors are as in Figs.~\ref{f_j11}---\ref{f_j05}.  From the accompanying animation, 
this event is largely homologous with jet~J18 (Fig.\,\ref{f_j18}), with the minifilament initially
residing on the far side of that lobe, and at the time of panels (a)---(d) it has crawled up to
the top of the lobe. In this case however, more of the filament remains near the solar surface 
than in the jet~J18 case, with some of the minifilament rotating onto the near side of the
lobe in panels (e)---(h) (blue arrows).  Thus, this is similar to the XRT-jet-causing minifilament 
eruption of Fig.\,\ref{f_j05} (at least for much of its life), with the minifilament largely remaining trapped in the base.  
Apparently it erupts and remains trapped in
the jet base, as depicted in Fig.\,\ref{f_schematic2}. Panel~(i) shows a zoomed-in XRT
view of the jet (axis ranges are different from those of the AIA panels);
the large dark oval, and similar smaller patches and spots, are artifacts.
The animation covers
2018 April 3, over approximately 16:25---16:55\,UT\@. 
The videos are constructed by performing a
running sum of every two consecutive images; the AIA videos are presented as full-cadence (12\,s)
running-sum movies, while the XRT video is presented as a running-sum movie of cadence $\sim$30---60\,s.  
The entire five-panel movie runs for $\sim$5\,s.}
\label{f_j20}
\end{figure}

Our second example in this category, shown in Figure~\ref{f_j20}, also displays a 
rolling (revolving) type motion.  This
jet,  jet~J20 of Table~1, occurs at the same location as jet~J18, occurring about 
100~min following that earlier jet (Jet~J18 is in Fig.~\ref{f_j18} and corresponding video, 
and Jet~J20 is in Fig.~\ref{f_j20} and corresponding video); thus jets~J18
and~J20 might be called homologous, although close inspection does reveal some
differences. In jet~J20, as with jet~J18, a minifilament erupts on the far side of
the M1---M2 (larger) magnetic lobe of the \citet{sterling.et15} schematic. It reaches the apex
of the lobe near the time of Figure~\ref{f_j20}(a---d).  Unlike jet~J18 however, very
little of the cool minifilament material obviously erupts outward; compare the frames in the
video in Figure~\ref{f_j18} near 15:14~UT, with those of the video of Figure~\ref{f_j20} near 16:43~UT;
relatively little of the cool material is erupting outward in the jet~J20 case,
compared to jet~J18.  On the other hand, more of the cool material continues moving
to the near side of the magnetic lobe than in the jet~J18 case, with the blue arrows
of Figure~\ref{f_j20}(e) showing prominent clumps of material on the near side of
that lobe.   Similar features can be seen in the jet~J18 video, but they are much
less apparent.  Thus apparently more of the minifilament was ejected in the jet~J18
case compared to the jet~J20 case.

We believe that jets~J20 (Fig.\,\ref{f_j20}) and J5 (Fig.\,\ref{f_j05}, at least for much
of its life) are showing
similar  behavior of EMFs, but from differing perspectives. 
Figure~\ref{f_schematic2} depicts a schematic corresponding to
that of \citet{sterling.et15}, showing these cases where the EMF is
mostly confined to the base region. In this case, the minifilament eruption
proceeds as in the \citet{sterling.et15} schematic through panel~(b), which is
Figure~\ref{f_schematic2} panel~(a).  Then however, the external open field encountered 
at the
apex of the M1---M2 magnetic lobe does not impede fully the movement of the
minifilament flux rope from right to left in the figure. It continues to move onto the far side
of the lobe.  As it does so, the magnetic field surrounding the minifilament/flux
rope continues to reconnect with the ambient coronal field.  This produces a hot
spire occurring progressively further from the eruption site and the JBP\@.  A cool
spire however (expected to be visible in AIA 304~\AA\ images) would be minimal or
absent, as long as  the external reconnection does not eat far enough into the
minifilament flux rope's core region.

In addition to events~J5 and~J20, other jets in our table have
partially confined EMFs.  The EMF in
jet~J16 is also clearly of this revolving type; it is seen from a perspective similar
to that of jet~J5.  Jet~J17 has an EMF that is somewhat harder to
see than the others.  It seems to be similar to these others, but the EMF appears to 
make it only partially past the apex of its M1---M2 lobe.
Thus, in general, these revolving type of EMFs follow the schematic picture of
Figure~\ref{f_schematic2}.  

Jet~J15, displayed in Figure~\ref{f_j15} and in the accompanying video, also  
perhaps has a confined EMF\@.  Over about 12:49---13:02~UT, the
minifilament appears to be erupting nearly radially from the surface.  Then, between about
13:02---13:12~UT, it moves first laterally toward the east, and then downward also,
appearing to return to the surface.  This is what is more commonly  regarded as a
confined filament eruption, whereby the eruption leaves the base region but then is
aborted, presumably by overlying strong-enough magnetic field
\citep[e.g.][]{moore.et01,ji.et03,sterling.et11a}.  Thus in this case, the confined
eruption basically follows the schematic of \citet{sterling.et15}, but where the
erupting cool minifilament material, along with much of the magnetic flux rope
surrounding and accompanying that cool material, has its outward eruption impeded by
overlying stronger field shortly after the situation depicted in panel~(c) of 
\citet{sterling.et15}.  If viewed from a different perspective (looking ``down 
from above"), this minifilament might also appear to be undergoing a revolving-type motion.

\section{Erupting Minifilament Characteristics}
\label{sec-characteristics}



\subsection{Erupting-Minifilament Physical Characteristics}
\label{sec-physical}

We have measured the lengths (projected on the plane of the sky) of the EMFs for each 
of the events, with
the results presented in Table~1.  There are several considerations in these
measurements:  the EMFs are sometimes difficult to discern, and we
cannot always  measure them at the same stage in their eruption (e.g., some are
visible at low altitudes, while others are obscured by  foreground material at low
altitudes). Also, they change shape during the  course of their eruption, and the
EMFs for different jets are not consistently visible in the same
AIA wavelengths (see below).  Despite these issues, the features we observe are
essentially always features showing in absorption in the 171, 193, and/or 211 EUV
channels (304 can have a strong  emission component), and therefore we expect that we
are seeing a substantial portion of the cool-material minifilament in each case.  In
other words, we expect that we are observing and measuring features representative of
the erupting cool-material minifilament for each case.  Given  these and other
factors however, our values should be considered as estimates good to within about a
few tens of percent.  With these caveats, the average values and standard deviations
for our measured EMF lengths are $14'' \pm 7''$.  The comments
column of Table~1 gives the specific AIA channel used for the measurement for each
case.

We also estimated velocities of the EMFs, by estimating the
distance traveled by the structure over a selected time range where we could track
the feature with fair confidence. Again there are various uncertainties that were
unavoidable. Perhaps the biggest was that we could not consistently observe the
erupting feature reliably during the same stage of eruption for differing events. 
For example, from on-disk observations, we have determined that the EMFs frequently 
start their  eruption with a relatively slow rise, and later
undergo an acceleration to a faster speed \citep{panesar.et20a}, similar to what is
frequently observed in erupting large-scale filaments 
\citep[e.g.,][]{tandberg.et80,sterling.et03,sterling.et07a,imada.et14,mccauley.et15,cheng.et20}.  
For our near-limb observations here however, it can be difficult to discern whether a
moving minifilament is in its slow-rise or fast-rise phase, because the initiation of
its rise is sometimes hidden from view by foreground chromospheric obscuring
material.  Also, the inability to determine reliably the orientation of the EMF 
with the line of sight makes speed determinations more complex.  (On the
other hand, the near-limb perspective allows for a better estimate of the radial
velocity, based on visual displacement over time,  of the EMF than
do the on-disk measurement, since the latter will have a  larger component of the
radial velocity directed along the line of sight)  Therefore, we again can only
present our measurements as an estimate of the rise speeds of the X-ray-jet-producing
EMFs during  the early stages of their eruption.  From all of the
results given in Table~1, we find  the average value and standard deviation of the
speeds to be $26\pm 13$\,\kms.  For each jet, we  estimated the speeds using the same
AIA channel as we used for measuring the EMF's size.

We have also measured the duration times for the jets, using the XRT images. This was
most straight forward, because these images were used for the original identification
of the jets.  Moreover, the visibility of the jet spires in the AIA images varied,
with some jets showing little or no spire in some of the AIA channels, despite having
an obvious spire in XRT\@.  We used as the duration the time over which the spire was
discernible in the accompanying XRT videos (Figs.~\ref{f_xrt1}---\ref{f_xrt4}). 
Usually some base brightening preceded the spire's appearance, and also
sometimes brightening happened after the spire's  disappearance; we ignored these base 
brightenings in
these duration estimates.  We estimate that there is an uncertainty of approximately
one minute in the durations given in Table~1.  From those values, we obtain an
average duration and standard deviation  of $21 \pm 10$\,min for our 21 events.

\subsection{Erupting-Minifilament Visibility}
\label{sec-visibility}

We can make an assessment of how readily visible the EMFs are for
our 21 events.  Because this is a visual assessment, there necessarily is a
subjective element to this exercise.  Moreover, the assessment   also depends upon
the manner in  which the data are examined; we have described our display methods for
the movies we use earlier: a running sum over two consecutive images, and  displayed
on a logarithmic scale after taking the fourth root of the intensities.  We do not
believe however that our assessment of whether an EMF was visible
would have changed substantially with a standard scaling of unsummed AIA images, on a
logarithmic scale but no further enhancements.  Most important was to zoom into the
eruption region close  enough, and to adjust the scaling maximum and minimum range
values appropriately.  With the realization however that the EMFs 
might be confined to the base region of the jet and possibly with little or no
obvious outward ejection into the corona, as with several examples in
\S\ref{sec-rolling}, and  as depicted in Figure~\ref{f_schematic2}, then the
identification can become  more apparent for some cases with the zoom-ins and 
maximum/minimum intensity scalings.

Of the 21 events, we assess that 11 of the X-ray jets have relatively
easily identifiable EMFs.  These are event numbers J2, J4, J5, J11,
J13, J14, J15, J16, J18, J20, and J21.  Several of these are displayed as examples in
\S\ref{sec-emf}.

We assess that six events have EMFs that are medium difficult to identify: J1, J6, J8, J9, J17, J19.
We assess that in three events an EMF is detectable but difficult to identify: J3, J7, J10.

In one event in particular, J12, it is difficult to identify unambiguously 
an EMF\@.  That event, however, is consistent with being driven by a minifilament-eruption
type of eruption.  This event is very small, with a base size of $\sim$15$''$, which complicates
possible EMF detection.  Moreover, the  location of
that jet's origin seems to be substantially obscured by absorbing foreground
structures, due to the jet's near-limb location. Activity that we see in AIA at the
jet's base appears to be consistent with a minifilament-eruption type of eruption.  Moreover, 
we can identify a candidate for an EMF, especially in 211\,\AA, over 
14:03---14:05~UT, but  the evidence is less compelling than
the other jets of this particular data set.  Thus we can say that  
we lack clear visual
evidence of an EMF in this case.

We do not know how the events having hard-to-see EMFs operate. 
If they do 
operate essentially the same as depicted in the schematics of 
\citet{sterling.et15} and Figure~\ref{f_schematic2}, then EMFs
might be hard to discern due to the events' small size and/or the diffuse coronal 
background. Their sizes are comparatively small; for example, we
estimated the sizes of J6, J7,  J8, and J9 as $\ltsim$10$''$, and so substantially
below the average value of $14''$. In some cases they are very diffuse and
ill-defined, as with jet~J19, and also jet~J1; in these cases what we see as a potential 
EMFs appear loosely consolidated, akin to the very large filament observed
by \hinode\ by \citet{berger.et11}, and it makes it hard to determine with certainty whether
the J1 minifilament was actually ejective or mainly confined (hence we say its ejection
is ``uncertain").  We also find that foreground ``coronal haze,''
resulting from either the jet occurring inside of the quiet Sun, or due to quiet Sun
corona being positioned in front of the jet along the line of sight, can make it
substantially more difficult to observe EMFs and jet spires, if they 
exist in these cases.

At other times, the EMFs appear to be present, but intrinsically faint.
Figure~\ref{f_j03} shows an example of this in jet~J3, where a likely EMF 
is faintly visible rising from about 12:50~UT to 13:04~UT, and then 
expanding outward as the jet spire forms over the next several minutes.  It is 
unclear whether this EMF is confined or ejective.  In 171 it 
appears as if part of it may be ejecting outward over about 13:10---13:20~UT, but
there is not a clear corresponding feature visible in the other three AIA 
channels.  Thus we cannot determine whether the minifilament truly ejects but this
is only apparent in 171, or if the apparent 171 feature is darkening due to 
field opening or some other factor.  All four AIA channels do however show some 
eruptive minifilament activity corresponding to the X-ray jet, but it is unclear
whether that EMF is ejective or confined in this cases.

\section{Standard (Narrow) and Blowout (Wide) Jets in the Set}
\label{sec-s_and_b}

We made an assessment of whether jets were standard or blowout based solely on the
appearance of their spires as wide or narrow in XRT images, which was a key assessment
consideration used by \citet{moore.et10,moore.et13}.  We call the XRT-narrow-spire jets
``standard" the XRT-wide-spire jets ``blowout," and ambiguous and/or uncertain jets we
labeled as ``uncertain."  Column~5 of Table~1 gives these assessments for each of our
events.   We identify 11 jets as blowout, four as standard, and six as uncertain.

Table~2 groups the jets based on the type: blowout, standard, or uncertain.  Of the
eleven blowout jets, only three came from clearly ejective minifilament eruptions.
Two of them appear to be from mainly confined eruptions, and another five are from either 
partially confined eruptions or from eruptions that were uncertain regarding whether the
cool minifilament material was ejected outward along the spire's open coronal field.

The four jets that we classify as standard, on the other hand, all originate from
eruptions of minifilaments that are either confined, partially confined, or
uncertain. That is, no clearly ejective EMF occurred in the four standard
jets, which, however, is a small sample of cases.

The jets that were classified as ``uncertain" regarding being standard or blowout 
had various spire appearances, as
described in the table.  These uncertain cases included both confined and
ejective minifilament eruptions.


\section{Summary and Discussion}
\label{sec-discussion}

We selected 21 X-ray solar polar coronal jets using \hinode/XRT images, and 
examined their origin using \sdo/AIA images.  Essentially all of these events show
evidence for originating from the eruption of minifilaments, consistent with the findings
of \citet{sterling.et15}. The ease of
detecting the EMF varied.  

Our set included 11 so-called blowout jets, if defined as jets whose spires grow to
be wide (comparable in size to the size of the jet's base) when viewed in XRT
images.  We found that these blowout jets can have EMFs
that appear to be ejected or confined, or where the eruption type is more
ambiguous.  Thus it is not only ejective EMFs that occur in 
blowout jets, as previously speculated \citep[e.g.][]{sterling.et15}.    Our
selection of 21 events only included four so-called standard jets, where the spire
remains narrow (compared to the jet's base) in XRT images.  None of these four cases
had EMFs that were clearly ejected, but instead were
confined, partially confined, or the nature of the minifilament eruption was
uncertain. This sample may be too small to conclude that clearly ejective
minifilament eruptions never make standard (narrow-spire) X-ray jets.

\begin{figure} 
\hspace*{-0.80cm}\includegraphics[angle=0,scale=0.7]{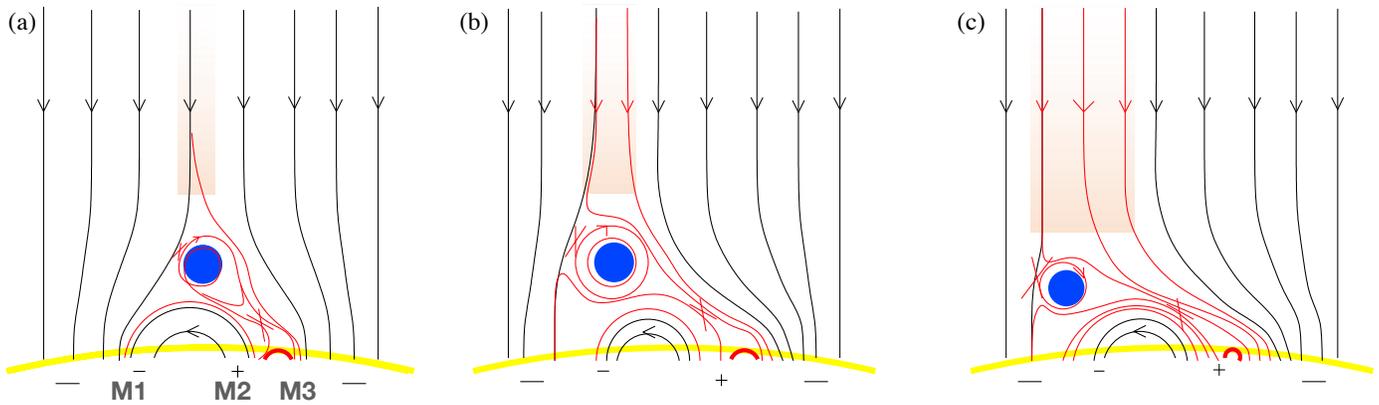} 
\caption{Schematic
picture of how some wide-spire jets work according to the minifilament eruption model, where 
the jet is made by a largely confined
minifilament eruption. (These would be called ``blowout jets" based on the \citeauthor{moore.et10}~\citeyear{moore.et10} and \citeyear{moore.et13} criteria based on the
broad X-ray spire, despite 
the EMF's flux rope being largely confined.)  Panel~(a) is the same as panel~(b) in Fig.~1 of \citet{sterling.et18}.  
Here, panels~(a), (b), and (c) together depict that the minifilament flux rope largely passes over the apex of the M1---M2 magnetic lobe, with a minimum amount of
the cool material escaping into the jet spire, much less than it is doing in the schematic of panel~(c) of \citet{sterling.et15}. 
Each red field line indicates a newly reconnected field line.  The shaded region widens 
along with the moving minifilament at the base, widening with
time away from from the JBP of the M2---M3 lobe; this occurs as the outside of the confined 
EMF's flux rope progressively
reconnects with field lines to the left in this perspective. } 
\label{f_schematic2} 
\end{figure}

\begin{figure}
\hspace*{0.0cm}\includegraphics[angle=0,scale=0.7]{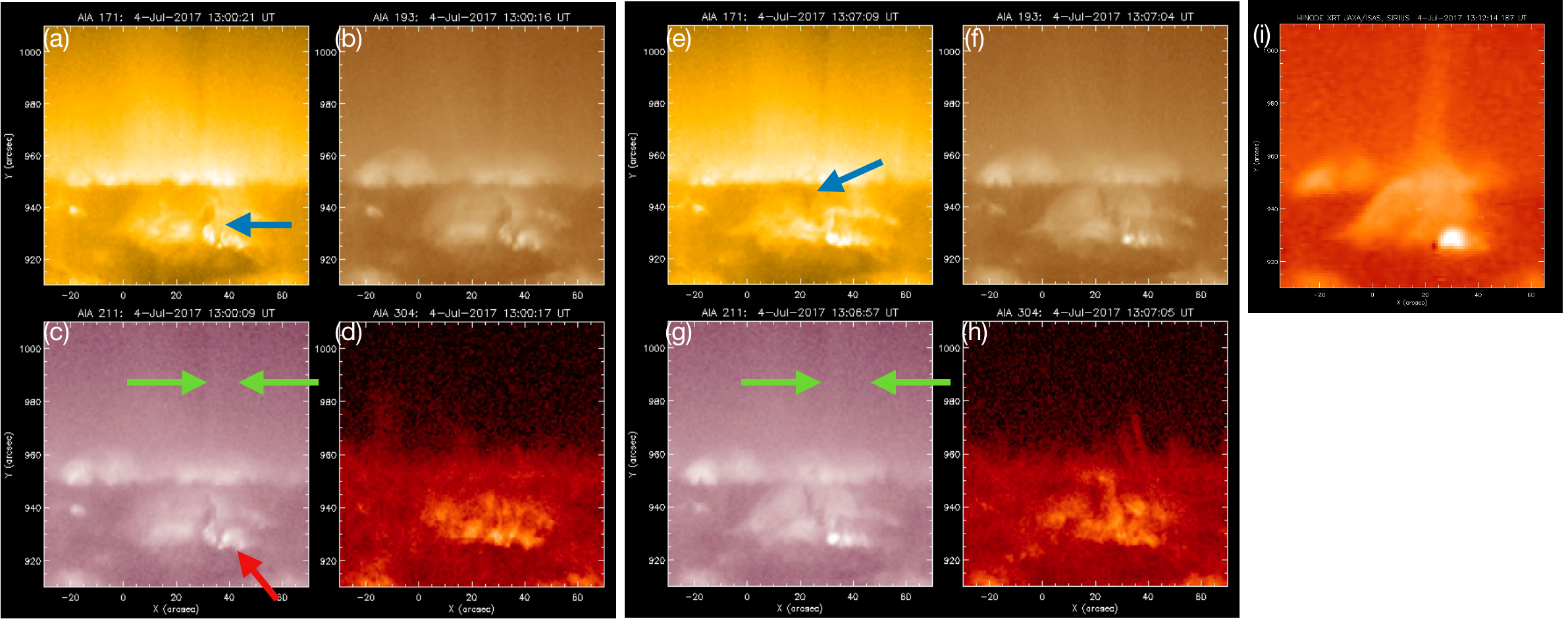}
\caption{
AIA images and an XRT closeup of Table~1 jet~J15.  The layout is as in 
Figs.~\ref{f_j11}---\ref{f_j20}, but for jet event~J15,
near 13:00~UT (left four panels) and near 13:07~UT (panels (e)---(h)), on 2017 July~4.
Arrow colors are as in Figs.~\ref{f_j11}---\ref{f_j20}. This is an example of a confined minifilament 
eruption of a different nature than those of Figs.~\ref{f_j05} and~\ref{f_j20}.  In those 
previous cases the EMF's flux rope had only enough energy to stay close 
to (ride on top of) the 
magnetic lobe M1---M2 of Fig.\,\ref{f_schematic2} (although in the 
case of Fig.~\ref{f_j05} we cannot determine whether the minifilament ejects outward later
in its evolution).  In this case, the EMF breaks 
free of the base region and starts to erupt into the higher corona, as in panels~(b) and (c) of
the schematic in \citet{sterling.et15}.
But its outward movement is largely stopped, presumably due to strong confining 
enveloping and surrounding 
 field,
and then returns to the surface.  This has been observed in confined large-scale filament eruptions
\citep[e.g.,][]{ji.et03,sterling.et11a}.  Panel~(i) shows a zoomed-in XRT
view of the jet (axis ranges are different from those of the AIA panels).
The animation covers
2017 July 4, over approximately 12:41---13:30\,UT\@. 
The videos are constructed by performing a
running sum of every two consecutive images; the AIA videos are presented as full-cadence (12\,s)
running-sum movies, while the XRT video is presented as a running-sum movie of cadence $\sim$30\,s.  
The entire five-panel movie runs for $\sim$5\,s.}
\label{f_j15}
\end{figure}

\begin{figure}
\hspace*{0.0cm}\includegraphics[angle=0,scale=0.7]{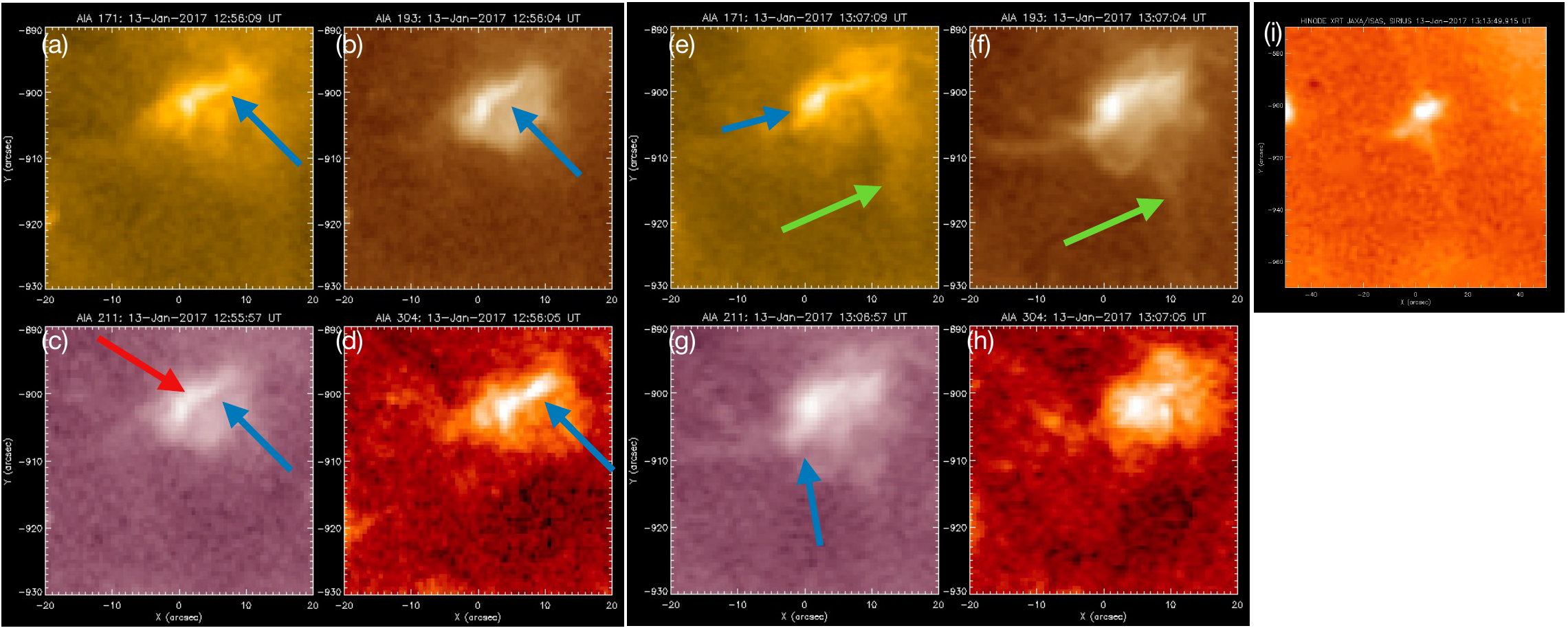}
\caption{
AIA images and an XRT closeup of Table~1 jet~J3.  The layout is as in Figs.~\ref{f_j11}---\ref{f_j20}, and Fig.\,\ref{f_j15}, but for jet event~J3,
near 12:56~UT (left four panels) and near 13:07~UT (right four panels, on 2017 January~13.
Arrow colors are as in Figs.~\ref{f_j11}---\ref{f_j20} and \ref{f_j15}. This is an example of a case of a jet-base EMF that we judged to be 
``difficult to identify."  
From the accompanying animation, we strongly suspect the feature indicated by the 
blue arrows is a faint EMF\@.  (Part of that EMF appears perhaps to be ejective, near the location of the green arrow in (e), but this cannot be confirmed in the other channels; see text.)  Fundamentally it follows the pattern of the schematic in \citet{sterling.et15} and 
\citet{sterling.et18}, where our perspective here is viewing that schematic from the right-hand
side; that is, we are seeing the minifilament face-on as it rises over the M1---M2 magnetic lobe
in the \citet{sterling.et18} schematic.  In the video, the EMF reaches near 
the lobe's apex at 
about 13:00~UT, and then it starts to expand in the SE---NW direction until about 13:20~UT, presumably 
as it reconnects with surrounding coronal field as depicted in the basic schematic of \citet{sterling.et15}.  In this case, the cool minifilament does not strongly continue to erupt radially from the Sun, as pictured in the \citet{sterling.et15} schematic, but remains partially trapped in the base region of the jet, as schematically depicted in Fig.\,\ref{f_schematic2}.  Panel~(i) shows a zoomed-in XRT
view of the jet (axis ranges are different from those of the AIA panels).
The animation covers
2017 January 13, over approximately 12:39---13:59\,UT for AIA, and over approximately 12:57---13:50\,UT
for XRT\@. 
The videos are constructed by performing a
running sum of every two consecutive images; the AIA videos are presented as full-cadence (12\,s)
running-sum movies, while the XRT video is presented as a running-sum movie of cadence $\sim$60\,s.  
The entire five-panel movie runs for $\sim$10\,s.}
\label{f_j03}
\end{figure}

For the confined-eruption case, we have shown clear cases where much of the minifilament
travels a substantial distance more-or-less horizontal to the solar surface, as
depicted schematically in Figure~\ref{f_schematic2}, rather then becoming trapped
near the apex of the base structure.

We can offer some speculations about why a largely confined minifilament eruption can 
still produce a broad-X-ray-spire ``blowout" jet.  One possibility is that the envelope surrounding
the confined cool minifilament undergoes extensive external reconnection with 
open field as the minifilament-carrying flux rope is traveling across the large base lobe, basically as depicted 
in Figure~\ref{f_schematic2}, where the reconnections persist more-or-less 
continuously as the EMF traverses the lobe from M2 to M1.  This could
occur, e.g, in a 3D situation whereby part of the EMF-carrying 
flux rope extends back toward the JBP, so that the external reconnections continuously
occur from the leading edge of the EMF back to the JBP, making 
a broad curtain-like spire.  This would therefore appear as a (broad-spire) blowout
jet.  When the EMF does not have continuous extensions backward 
toward the JBP, then it could be that only a short leading edge of the EMF 
undergoes external reconnection, forming a (narrow-spire) standard jet.  Another 
possibility for a confined EMF 
to occur with a (broad-spire) blowout jet is that there could be a flux rope without 
cool material 
situated above the cool-material minifilament-carrying flux rope represented 
by the blue circle in Figure~\ref{f_schematic2}.  That higher-altitude flux rope
would be essentially invisible in our X-ray and EUV images, but it could be 
destabilized by the eruption of the cool-material-carrying minifilament flux rope,
and undergo an ejective eruption that makes the broad-spire jet.  This would be similar to 
the ``lid-removal''-type mechanism occurring in some large-scale filament eruptions
\citep{torok.et11,lynch.et13,sterling.et14}, as in double-decker (or multiple-decker) 
filament eruptions whereby one or more filaments stacked at different heights above
the same neutral line sequentially erupt \citep[e.g.][]{su.et11,joshi.et20}.  Which
mechanism is acting perhaps could be a factor in whether the spire resulting from
the confined minifilament eruption becomes
wide, as depicted in Figure~\ref{f_schematic2}, thus forming a blowout jet, or 
remains narrow, thus forming a standard jet.  Details of
whether these specific mechanisms are operational in our events is beyond the 
scope of the present study.

Of the 21 the events, for 11 of them, or about 50\%, we assess that the 
identification of the EMFs was relatively easy, using standard
high-cadence AIA videos focused in on the jet-base region (usually with a FOV of
$100'' \times 100''$  or smaller).  In nine of the remaining cases, about 40\%, we
saw evidence for  minifilament eruptions that we classified as medium-difficult or
difficult to identify. These non-easy cases were jets where we were still confident
that a minifilament eruption was occurring, but which are more difficult to make out visually.
In only one case were we not confident of discerning at least a faint EMF\@.

Regarding visibility of EMFs in different AIA channels, corresponding to X-ray
jets:  In some cases, the EMFs 
accompanying our X-ray jets were visible in as few as one of the four EUV channels we 
examined, and frequently they were not
equally apparent in all four channels. Therefore it is critical to examine multiple
EUV wavelengths before concluding that any given jet may not have an
EMF at its base. This is similar to the situation with X-ray jet 
observations prior to availability of high-cadence EUV images: EMFs are essentially
invisible in soft X-rays; to see them, one must also inspect concurrent EUV images.
Here we have learned that one cannot always rely on a single EUV wavelength channel
for a complete search of the EMFs that might occur at the base of a given X-ray-observed
jet.

The lengths of the EMFs are $14'' \pm 7''$.  This is not far
different from the values of $11'' \pm 7''$ found by \citet{sterling.et15}, who also
looked at minifilaments that produced polar coronal hole XRT jets.  These values are
somewhat smaller than the $\sim$$25''$ average minifilament size, and $\sim$23$''$
for jet-base widths, reported for ten  quiet-Sun jets by \citet{panesar.et16a}.  That
\citet{panesar.et16a} study, however, estimated the full (curved) length of the
minifilament viewed on disk prior to its eruption; we instead measure the length of
the minifilaments at the start of or just after eruption onset, and we measure the
projected length against the plane of sky, without attempting to estimate the full
length along the axis of the minifilament.  Another difference is that the
\citet{panesar.et16a} jets were selected from full-disk AIA images, while we used XRT
images focused in on a smaller FOV; it is unclear how jets selected from AIA compare
with those selected via our procedure.   \citet{panesar.et18a} measured the base size
of 13 AIA-observed on-disk coronal hole jets, and found values of $16''\pm 8''$, where they used
the methods of \citet{panesar.et16a} to find the base size of those coronal-hole jets.  
(One might assume that, similar to what was found in
\citeauthor{panesar.et16a}~\citeyear{panesar.et16a}, the lengths of the minifilaments
of \citeauthor{panesar.et18a}~\citeyear{panesar.et18a} pre-eruption minifilaments are
probably similar to the size of the jet-base region.)

We found speeds of the EMF, when they were in the process of erupting, to be
$26\pm 13$\,\kms.  This is substantially faster than the values near $1$\,\kms\ reported by
\citet{panesar.et20a} from the earliest stages of the onset of their minifilament eruptions.  
Here we are likely favoring speeds of the minifilaments in their fast-rise phase, while those 
of \citet{panesar.et20a} are favoring the slow-rise stages.  There may also be a difference
due to the above-mentioned viewing perspective, with our observations here showing the rising 
minifilaments in profile (so that radial outward motions would show as a displacement in the 
plane of the sky),while 
the \citet{panesar.et20a} study would, on average, have a larger component of the radial 
velocity along the line of sight.  Thus the \citet{panesar.et20a} investigation may have 
systematically measured somewhat lower velocities than our study.

\cite{kumar.et19} examined 27 on-disk coronal-hole jets using AIA 304, 171, and 193\,\AA\
images, and report that 67\% ``are associated with" minifilament ejections/eruptions. They also
report that the remaining events also displayed evidence for an eruption of a filament channel
without cool material.  These findings are in general agreement with what we find here, and in
the studies referenced in the preceding paragraphs.  Nonetheless, in light  of our results
presented here \citep[and also discussed in][]{sterling.et15}, we wonder whether some of those
remaining 33\% in the \citet{kumar.et19} study might have EMFs that erupted in the manner 
described in Figure~\ref{f_schematic2}, and thus were not recognized as minifilament eruptions
in that study because of the lack of a clear minifilament ejection into the higher corona.

\begin{acknowledgments}
The authors received funding from the Heliophysics Division of NASA's Science 
Mission Directorate through the Heliophysics Guest Investigators (HGI) Program, and 
the MSFC \hinode\ Project. \hinode\ is a Japanese mission developed and launched 
by ISAS/JAXA, with NAOJ as domestic partner and NASA and UKSA as international 
partners. It is operated by these agencies in co-operation with ESA and NSC (Norway). 
We acknowledge the use of AIA data. AIA is an instrument onboard \sdo, a mission of
NASA's Living With a Star program.
\end{acknowledgments}

\bibliography{ms_arxiv_sterling_et_220128}
\clearpage


\startlongtable
\begin{longrotatetable}
\begin{deluxetable}{ccccccccc}
\tablewidth{3cm}
\tabletypesize{\scriptsize}

\tablecaption{Examined Coronal Jets \label{table1}}
\tablehead{\colhead{Event} &\colhead{Date} & \colhead{Start} &\colhead{Duration} & \colhead{Type} & \colhead{Clear MF ejection?} & 
\colhead{Velocity [\kms]} & \colhead{MF Length (arcsec)}  & Notes \\
} 
\startdata
J1 & 2017 Jan 13 & 12:30 & 27 & Blowout & Uncertain & 14  & 10---15 & \parbox{5cm}{\vspace{0.3cm} Swirling amorphous MF
expelled from JBP location with substantial horizontal movement, perhaps best visible in 304.  MF material may flow out
along open field, but cannot confirm this, hence it is uncertain whether the EMF is ejective or (partially) confined. \vspace{0.3cm}}\\
J2 & 2017 Jan 13 & 12:57 & 07 & Uncertain & Yes & 37  & $\sim$7 & \parbox{5cm}{\vspace{0.3cm} Measured in 171. EMF clearly appears as an erupting loop in all four channels.  Measured size is of whole minifilament in 171 at 12:54\,UT  ($\sim$erupting loop footpoint separation distance).\vspace{0.3cm}}\\
J3 & 2017 Jan 13 & 13:02 & 39 & Uncertain & Uncertain & 12  & $\sim$12 & \parbox{5cm}{\vspace{0.3cm} A weak jet in XRT and relatively small in size, making details hard to assess. MF eruption appears at least partially ejective and perhaps partially confined in 171, but appears to be largely confined in the other three channels. Velocity from 193; size from 193 at 12:55\,UT\@. \vspace{0.3cm}}\\
J4 & 2017 Jan 13 & 13:42 & $\ge$14& Blowout & partial & 52  & $\sim$20 & \parbox{5cm}{\vspace{0.3cm} Measured in 171.  Size of whole minifilament in 171 at 13:42 is $\sim$20$''$, but probably only its western half erupts.\vspace{0.3cm}}\\
J5 & 2017 Mar 6 & 10:58 & 28 & Blowout & No & 23  & 25 & \parbox{5cm}{\vspace{0.3cm} Confined EMF, expelled from JBP location and confined to revolving motion over neighboring magnetic lobe, as in Fig.\,\ref{f_schematic2}; clear in all four channels.  Size of minifilament measured in 304 11:06:57.\vspace{0.3cm}}\\
J6 & 2017 Mar 6 & 11:48 & 21 & Standard & Partial ejection & 22  & $\sim$6---10 & \parbox{5cm}{\vspace{0.3cm} Erupting loop MF, with at least some of the material leaking onto vertical field near loop apex; visible in 
all but 304; hard to see in 304, perhaps due to small size and near-limb location.  Size is of EMF-loop base in 193 over $\sim$11:48---11:51.\vspace{0.3cm}}\\
J7 & 2017 Mar 6 & 11:53 & 15 & Uncertain & Faint EMF & 19  & $\sim$4---7 & \parbox{5cm}{\vspace{0.3cm} Maybe best in 211; too small and too near limb for 304.  Small blob moving horizontally, and may leak out of spire, but hard to tell.  Size is rough 
estimate of blob width in 211 at 11:53:45.\vspace{0.3cm}}\\
J8 & 2017 Mar 6 & 12:05 & 23 & Standard & No & 9  & $\sim$10 & \parbox{5cm}{\vspace{0.3cm} EMF appears as small feature moving horizontally at base of spire, maybe best seen in 193 and 211.  Size is rough 
estimate of length (perhaps foreshortened) in 193 at 12:10:09.\vspace{0.3cm}}\\
J9 & 2017 Mar 6 & 12:37 & 21 & Standard & Confined(?) & 54  & $\sim$12 & \parbox{5cm}{\vspace{0.3cm} Homologous with J6.  Clear erupting-loop minifilament eruption, visible in all channels, maybe best contrast 
in 211.  Size is rough 
estimate of projected length measured in 211 at 12:17:45.\vspace{0.3cm}}\\
J10 & 2017 Mar 6 & 12:48 & 15 & Blowout & Uncertain & 15---30  & $\sim$13 & \parbox{5cm}{\vspace{0.3cm} 
Very faint minifilament erupts outward in 193, but not really visible in other
channels; appears to ``leak out" from apex of bright base over about 12:48 to 12:56.  Measured size is that of the 
leaking-out structure in 193 at 12:51:09.\vspace{0.3cm}}\\
J11 & 2017 Mar 6 & 13:18 & 17 & Uncertain & First confined, then leaks out & 35  & $\sim$14 & \parbox{5cm}{\vspace{0.3cm} 
Clear EMF in all channels.  Becomes confined at ~13:19,
but then leaks out the spire until at least 13:44. Size measured in 171, just prior to leak out.\vspace{0.3cm}}\\
J12 & 2017 Mar 6 & 13:59 & 8 & Uncertain & No clear EMF & ---  & --- & \parbox{5cm}{\vspace{0.3cm} 
Faint jet in XRT, but very bright in AIA\@.  No clear EMF\@. A candidate is present, especially 
in 211\,\AA] over 14:03---14:05~UT, but difficult to separate unambiguously from other surrounding
outflows. Any potential EMF is likely largely hidden due to small size (bright base 
extent $\sim$12$''$ at 14:04:09~UT) and obscuration by spicule forest.\vspace{0.3cm}}\\
J13 & 2017 Mar 6 & 14:38 & 14 & Blowout & Yes & 48  & $\sim$13 & \parbox{5cm}{\vspace{0.3cm} 
Sigmoid-shaped EMF being ejected, clear in all channels sans 304.  MF seems to split 
upon eruption.  Size (linear extent) measured in 171 at 14:40:21.  Velocity is of erupting west lobe of sigmoid
filament over 14:40:57---14:41:57~UT\@.\vspace{0.3cm}}\\
J14 & 2017 Mar 6 & 15:42 & 8 & Blowout & Yes & 8; 22  & $\sim$8 & \parbox{5cm}{\vspace{0.3cm} Similar to 13.  Clear erupting sigmoid-shaped minifilament; visible 
in all channels.  MF seems to split upon eruption, and only part of it is ejected. 
Size (linear extent) measured in 211 at 15:37:45.  Slower velocity is pre-eruption expansion of
east lobe of sigmoid MF over  15:34:21 - 15:37:33; faster velocity 
is of erupting east lobe of sigmoid MF over 14:44:09 - 15:45:33~UT\@.\vspace{0.3cm}}\\
J15 & 2017 Jul 4 & 12:59 & $\>$33 & Blowout & No (aborted ejection) & 18  & $\sim$5 & \parbox{5cm}{\vspace{0.3cm} EMF that is clearly confined, apparent in all channels.  Velocity from 171 during eruption, over 12:55:21 - 13:02:09~UT\@.  Size is from 171 at 12:57:55~UT, 
but it is likely severely foreshortened. \vspace{0.3cm}}\\
J16 & 2017 Jul 4 & 16:07 & 19 & Standard & No (revolving) & 33  & $\sim$15 & \parbox{5cm}{\vspace{0.3cm} Clear 
confined EMF, best seen in 304.  Velocity from 304 during eruption, over 16:09:57 - 16:11:57~UT\@.  Size 
is ~$15''$ from 304 at 16:12:21~UT, likely severely foreshortened. \vspace{0.3cm}}\\
J17 & 2018 Apr 3 & 14:19 & 43 & Uncertain & Mainly confined (but perhaps leaks out) & 33  & $\sim$15 & \parbox{5cm}{\vspace{0.3cm} 
Confined EMF, but some cool material may leak out near apex of MF's eruption.  Best visible in 193 and 211 as short extended silhouette, more faintly
visible in 171; not apparent in 
304.\@.   Velocity is from 211 over 14:18:09 - 14:19:45~UT.  Size is from 211 at 14:18:33. \vspace{0.3cm}}\\
J18 & 2018 Apr 3 & 14:52 & 17 & Blowout & Partial ejection & 23, 26  & $\sim$18 & \parbox{5cm}{\vspace{0.3cm} 
EMF first confined, but then part of it ejecting outward.  Clear in all channels.
 Velocities and size from 171.  Slower velocity is MF moving up onto large lobe, faster velocity is that of an erupting
 strand.  Size of $\sim$18$''$ is from 15:04:57, when MF is seen in profile and so thus perhaps with minimal foreshortening. 
\vspace{0.3cm}}\\
J19 & 2018 Apr 3 & 15:23 & $\gtsim$37 & Blowout & Faint, and obscured by haze & $\ltsim$30  & $\sim$28 & \parbox{5cm}{\vspace{0.3cm} 
Weak X-ray jet.  Also hazy in AIA, but there is a clear EMF, and it is probably ejective, 
erupting over 15:32 - 15:57, weakly detectable in all channels. Seems 
to be a non-consolidated MF\@. Derived velocity, from 171, is from a fast-moving easily discernible 
substructure that perhaps moves faster than other parts of the MF\@.  Size is that of the base of the 
EMF structure in 171. 
\vspace{0.3cm}}\\
J20 & 2018 Apr 3 & 16:35 & 12 & Blowout & Mainly confined & $\sim$11-15  & $\sim$25 & \parbox{5cm}{\vspace{0.3cm} 
Homologous with J18. Same type of MF, and same type of "crawl up" is visible in all channels.  There 
is some partial ejection over 16:41-16:45~UT\@. Derived velocity from 171 is of MF moving up to apex
of large lobe.  Size is from 171 when it is near the apex of that lobe.
\vspace{0.3cm}}\\
J21 & 2018 Apr 3 & 16:58 & 22 & Blowout & Yes & $\sim$10-13  & $\sim$15---20 & \parbox{5cm}{\vspace{0.3cm} 
Homologous with J17.  Clear erupting feature visible in all channels, but best visible as 
a (subtle) dark EMF feature in 193 and 211.  Velocity is of pre-eruption MF slow rise from 193.  Size is length in 193,
perhaps seen in profile (and so reduced foreshortening).
\vspace{0.3cm}}\\
\enddata

\end{deluxetable}
\end{longrotatetable}

\begin{deluxetable}{ll@{\hskip 1.5cm}l}
\tabletypesize{\footnotesize}
\tablecaption{Blowout/Standard Jets, and Erupting Minifilaments \label{tab:table2}}
\tablehead{
\colhead{Jet} & \colhead{Type} &\colhead{EMF's Degree of Jet Base Confinement}
}
\startdata
J1 &	Blowout	 &	Uncertain \\
J4 &	Blowout	 &	Partial \\
J5 &	Blowout	 &	Confined \\
J10 &	Blowout	 &	Uncertain \\
J13 &	Blowout	 &	Ejective \\
J14 &	Blowout	 &	Ejective \\
J15 &	Blowout	 &	Confined \\
J18 &	Blowout	 &	Partial \\
J19 &	Blowout	 &	Uncertain \\
J20 &	Blowout	 &	Confined \\
J21 &	Blowout	 &	Ejective \\
\hline
\hline
J6 & Standard   & Partial \\
J8 & Standard   & Confined \\
J9 & Standard   & Uncertain \\
J16 & Standard   & Partial \\
\hline
\hline
J2 &	Uncertain \hspace{0.5cm}\parbox{4cm}{(Spire between narrow and broad.)} &     Ejective \\
J3 & 	Uncertain \hspace{0.5cm}\parbox{4cm}{(Borderline too small and weak for assessment.)}   & Uncertain \\
J7 &	Uncertain \hspace{0.5cm}\parbox{4cm}{(Narrow bright spire, with faint broader component.)} &     Ejective \\
J11 &   Uncertain \hspace{0.5cm}\parbox{4cm}{(Too faint for clear determination.)} &     Mixed \hspace{0.5cm}\parbox{4cm}{(First confined, then ejective.)}\\
J12 &   Uncertain \hspace{0.5cm}\parbox{4cm}{(Narrow bright spire, with possible faint broad component.)}&     None \hspace{0.5cm}\parbox{4cm}{(Suspected, but cannot be confirmed due to faintness and small size.)}\\
J17 &   Uncertain \hspace{0.5cm}\parbox{4cm}{(Spire between narrow and broad.)} &     Confined \\
\enddata
\end{deluxetable}
\clearpage

\clearpage

\clearpage

\clearpage

\clearpage

\clearpage

\clearpage

\clearpage

\clearpage

\clearpage

\clearpage

\clearpage

\end{document}